\documentclass[twocolumn,twocolappendix]{aastex631}

	\usepackage{graphicx}	
	\usepackage{amsmath}
	\usepackage{amssymb}	
	\usepackage{longtable}
	\usepackage{threeparttable}
	\usepackage{multirow}
	\usepackage{booktabs}
	\usepackage{array}
    \usepackage[figuresright]{rotating}
    \usepackage[T1]{fontenc}

	\shorttitle{HS 2325+8205}
	\shortauthors{Sun et al.}

	\graphicspath{{./}{figures/}}

\begin{document}
			\title{
				Tilted Disk Precession and Negative Superhumps in HS 2325+8205: A Multi-Window Analysis}


\correspondingauthor{Sheng-Bang Qian}
\email{qsb@ynao.ac.cn}	
\author{Qi-Bin Sun}
\affiliation{Department of Astronomy, School of Physics and Astronomy, Yunnan University, Kunming 650091, PR of China}
\affiliation{Key Laboratory of Astroparticle Physics of Yunnan Province, Yunnan University, Kunming 650091, PR China}

\author{Sheng-Bang Qian}

\affiliation{Department of Astronomy, School of Physics and Astronomy, Yunnan University, Kunming 650091, PR of China}
\affiliation{Key Laboratory of Astroparticle Physics of Yunnan Province, Yunnan University, Kunming 650091, PR China}

\author{Li-Ying Zhu}
\affiliation{Yunnan Observatories, Chinese Academy of Sciences, Kunming 650216, PR of China}
\affiliation{University of Chinese Academy of Sciences, No.19(A) Yuquan Road, Shijingshan District, Beijing, PR of China}

\author{Qin-Mei Li}
\affiliation{Department of Physics, college of physics, Guizhou university, Guiyang, 550025, PR of China}

\author{Min-Yu Li}
\affiliation{Yunnan Observatories, Chinese Academy of Sciences, Kunming 650216, PR of China}
\affiliation{University of Chinese Academy of Sciences, No.19(A) Yuquan Road, Shijingshan District, Beijing, PR of China}		

\author{Ping Li}
\affiliation{Yunnan Observatories, Chinese Academy of Sciences, Kunming 650216, PR of China}
\affiliation{University of Chinese Academy of Sciences, No.19(A) Yuquan Road, Shijingshan District, Beijing, PR of China}	
			

			\begin{abstract}

Tilted disk precession exists in different objects. Negative superhumps (NSHs) in cataclysmic variable stars (CVs) are believed to arise from the interaction between the reverse precession of a tilted disk and the streams from the secondary star. 
Utilizing TESS photometry, we present a comprehensive investigation into the tilted disk precession and NSHs in the dwarf nova (DN) HS 2325+8205, employing eclipse minima, eclipse depths, NSH frequencies, and NSH amplitudes and the correlation between them as the windows. We identified NSHs with a period of 0.185671(17) days in HS 2325+8205. 
The NSH frequency exhibits variability with a period of 3.943(9) days, akin to the tilted disk precession period validated in novae-like stars (NLs, SDSS J0812) and intermediate polars (IPs, TV Col).
The O-C of eclipse minima were similarly found to vary cyclically in period 4.135(5) days, characterized by a faster rise than fall. 
Furthermore, the NSH amplitude exhibits complex and diverse variations, which may be linked to changes in the disk radius, mass transfer rate, and the apparent area of the hot spot.
For the first time in DNe, we observe bi-periodic variations in eclipse depth ($\rm P_{1} = 4.131(4)$ days and $\rm P_{2} = 2.065(2) $ days $\rm \approx P_{prec}/2  $), resembling those seen in IPs, suggesting that variations with $\rm P_{2}$ are not attributable to an accretion curtain, as previously suspected.
Moreover, NSH amplitude and eclipse depth decrease with increasing NSH frequency, while NSH amplitude correlates positively with eclipse depth.
These complex variations observed across multiple observational windows provide substantial evidence for understanding of tilted disk precession and NSHs.

\end{abstract}



\section{Introduction} \label{sec:intro}

Superorbital signals (SORs) have been found in Cataclysmic variable stars (CVs) such as TV Col \citep{Sun2024ApJ} ($\rm P_{\text{sor}} \sim 4$ d), Active Galactic Nuclei, Black Hole Binaries (BHBs) like SS 433 \citep{Foulkes2010MNRAS} ($\rm P_{\text{sor}} \sim 162$ d), and X-ray Binaries (XRBs) such as Her X-1 \citep{Giacconi1973ApJ} ($\rm P_{\text{sor}} \sim 35$ d). These signals are generally believed to originate from the reverse precession of tilted accretion disks. Evidence comes from the discovery of reverse precession in jets (e.g., \citealp{Hjellming1981ApJ}; \citealp{Abraham2018NatAs}; \citealp{Caproni2004ApJ}; \citealp{2023NaturM87}). The precession period of the jet in M87 was measured to be $\rm P_{\text{prec}} \sim 11$ yr \citep{2023NaturM87}, and the precession timescale of the Active Galactic Nucleus 2MASXJ12032061+1319316 was estimated to be $\rm P_{\text{prec}} \sim 10^5$ yr \citep{2017MNRASAGN}.

CVs exhibit SORs with periods of only a few days, believed to originate from the reverse precession of tilted disks  \citep[e.g.,][]{Katz1973NPhS..246...87K, Barrett1988MNRAS, harvey1995superhumps, wood2009sph}. They have short precession timescales and numerous examples of precession signals \citep[e.g.,][]{2023MNRAS.520.3355S, sun2022study, Bruch2022, Bruch2023, BruchIII}. Therefore, CVs provide an ideal experimental environment for studying tilted disk precession in astrophysics.
Negative Superhumps (NSHs) in CVs, with periods several percent shorter than the orbital period, are thought to result from interactions between the streams from the secondary star and the reverse precession of tilted disks \citep[e.g.,][]{1985A&A...143..313B, patterson1999permanent, harvey1995superhumps}. The relationship $\rm 1/P_{\text{prec}} = 1/P_{\text{sor}} = 1/P_{\text{nsh}} - 1/P_{\text{orb}}$ exists, where $\rm P_{\text{prec}}$ is the precession period.

Evidence for the tilted disk precession has been found, with the surprising agreement of the SORs with predictions of the tilted disk precession period ($\rm P_{\text{prec}} \simeq P_{\text{sor}} $) being the most important evidence. Other evidences are the eclipse minima, the eclipse depth, the eclipse width, and the NSH amplitude with periodic variations similar to the theoretical tilted disk precession period ($\rm P_{\text{prec}} \simeq P_{\text{sor}} \simeq P_{\text{O-C, eclipse}} \simeq P_{\text{depth, eclipse}} \simeq P_{\text{width, eclipse}}\simeq P_{\text{amplitude, NSHs}}$). 
NSHs are thought to originate from the interaction of the streams with the tilted disk precession, and the main observational evidence for this is $\rm 1/P_{\text{nsh}} \simeq 1/P_{\text{sor}} + 1/P_{\text{orb}}$. Second, some authors have reproduced NSHs based on smooth particle hydrodynamics within the framework of this model (e.g., \citealp{wood2009sph}; \citealp{Montgomery2009MNRAS}; \citealp{2021PASJ...73.1225K}). The finding that the NSH amplitude varies as the tilted disk precession is very important evidence. However, these observational evidences are still lacking.

Recently, we conducted an extensive study on tilted disk precession in CVs using space telescope data, focusing on NSHs as a key indicator \citep[e.g.,][]{Sun2023arXiv, 2023arXiv230905891S, 2023arXiv230911033S, Sun2024ApJ}. DN outbursts are commonly attributed to thermal instabilities in accretion disks, while NSHs are believed to arise from interactions between tilted disk precession and streams from secondary stars. However, the combination of thermally unstable and tilted disks remains poorly understood. We observed in DN systems AH Her, ASAS J1420, TZ Per, and V392 Hya that NSH amplitudes vary with outburst phases \citep{2023arXiv230905891S, 2023arXiv230911033S}, suggesting a significant link between NSHs and DN outbursts as a valuable avenue for studying accretion disk instability and NSH origins.
Additionally, we detected periodic variations in eclipse minima, eclipse depth, and NSH amplitudes consistent with the precession period of the tilted accretion disk in the NL SDSS J0812 \citep{Sun2023arXiv}. These findings provide strong evidence for a direct correlation between NSH origin and tilted disk precession. Subsequently, we confirmed these phenomena in the Intermediate Polars (IPs) system TV Col, a prototype for NSH systems, and observed a bi-periodic variation in eclipse depth, potentially linked to the dual accretion curtains of IPs \citep{Sun2024ApJ}, offering critical observational insights into tilted disk precession in CVs.
Recent studies by \cite{Bruch2024ApJS} have similarly identified a prevalence of periodic variations in O-C diagrams and eclipse widths, consistent with the period of tilted disk precession.

HS 2325+8205, a long-period eclipsing DN ($\sim$ 4.66 hr), initially discovered by \cite{morgenroth193677}, has a rich observational history \citep[e.g.,][]{morgenroth193677, hagen1995hamburg, aungwerojwit2005hs, shears2011hs2325+, hardy2017hunting, Sun2023MNRAS}, and is classified as a Z Cam-type DN due to its short recurrence period \citep{pyrzas2012hs}. In our recent study, we identified quasi-periodic oscillations (QPOs) with a period of $\sim$ 2160 s during the peak of a long outburst, with the QPO intensity correlated with orbital phase \citep{Sun2023MNRAS}. Hence, we propose that studying the relationship between QPOs and orbital phase could offer valuable insights into their origins.

Although the tilted disk reverse precession model is widely used, the mechanism that leads to disk tilting and reverse precession is controversial. Fundamental questions, such as how the material streams from the secondary star interact with the accretion disk and the underlying physical processes at the point of contact, require more detailed exploration.
SORs characterize tilted disk precession, but current studies indicate that SORs are not universally observed in systems with NSHs, particularly in DNe with NSHs \citep[e.g.,][]{ramsay2017v729, court2020ex, 2023arXiv230905891S, 2023arXiv230911033S}, posing a challenge to the tilted disk precession theory. HS 2325+8205, a DN with NSHs, presents new observational data. 
Utilizing eclipse minima, eclipse depths, NSH frequencies, and NSH amplitudes as investigative windows, we endeavor to explore potential evidence of tilted disk precession and validate whether phenomena observed in Non-magnetic CVs (NLs) and Magnetic CVs (IPs) also manifest in DNe. This approach aims to provide invaluable observational insights for studies pertaining to tilted disk precession and NSHs.

The organization of this paper is outlined as follows:
Section \ref{sec:style} introduces the photometric data obtained through the TESS mission.
In Section \ref{sec:Analysis}, we conduct an in-depth analysis of NSH frequency, NSH amplitude, eclipse depth, as well as O-C variations and QPOs.
Section \ref{sec:DISCUSSION} delves into potential explanations for the observed changes.
Finally, Section \ref{sec:CONCLUSIONS} presents the conclusions.

\begin{figure}
\centerline{\includegraphics[width=\columnwidth]{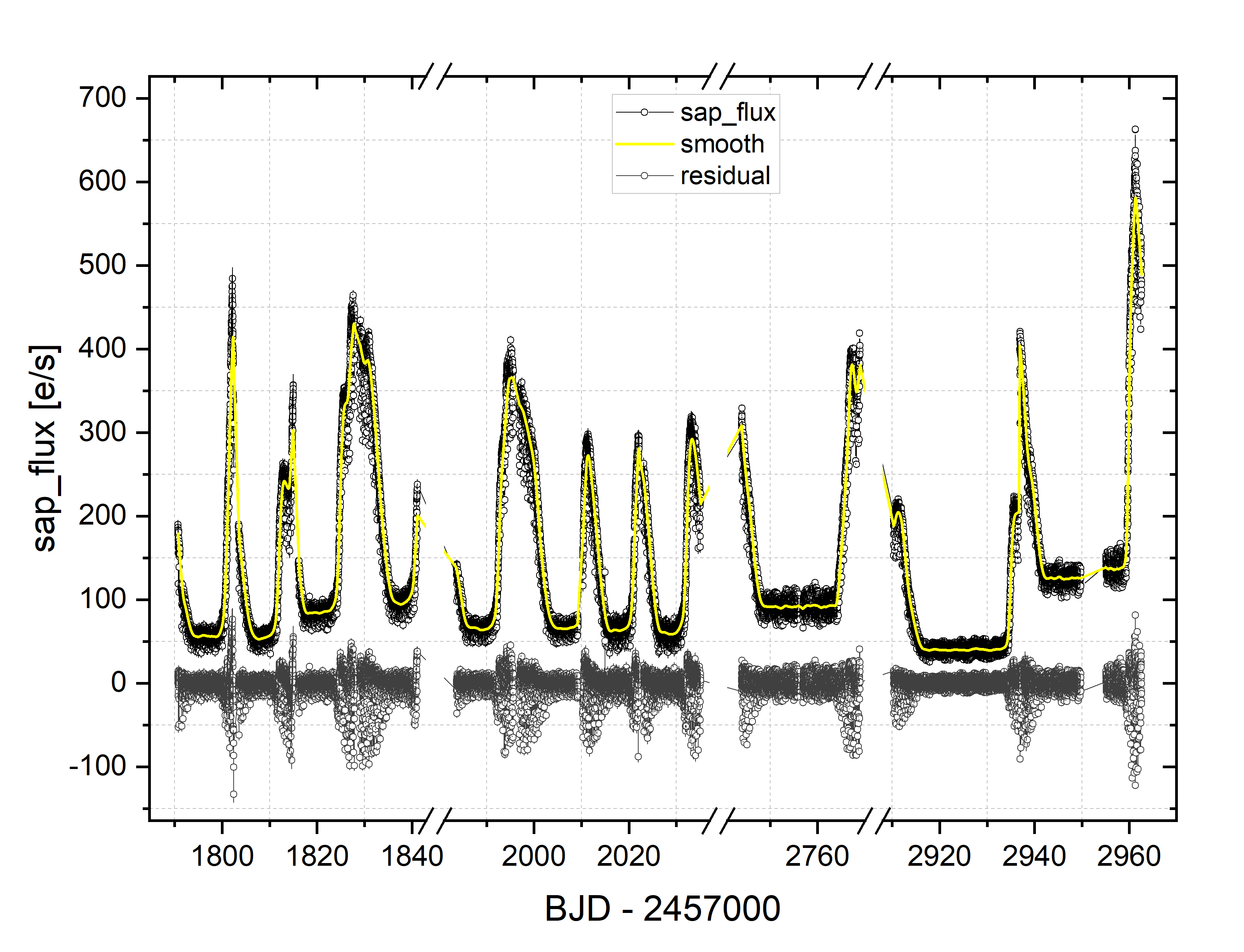}}
\caption{HS 2325+8205 light curve from TESS photometry. The yellow line is the LOWESS fit in order to remove the outburst trend.\label{light}}
\end{figure}


\section{TESS Photometry} \label{sec:style}

\begin{deluxetable}{llll}
	\tablecaption{Journal of Observations.\label{LOG}}
	\tablewidth{0.01pt}
	\tabletypesize{\scriptsize}
	\tablehead{	
		\colhead{Sectors}&\colhead{ Start time}&\colhead{ Start time}&\colhead{End time}\\
			\colhead{ }&\colhead{ MBJD}&\colhead{ UT}&\colhead{MBJD }
	}
	\startdata
18&58790.15713&2019-11-03&58814.51574\\	
19&58815.58519&2019-11-28&58840.65031\\	
25&58983.12924&2020-05-14&59008.80275\\	
26&59009.76386&2020-06-09&59034.63223\\	
53&59743.49271&2022-06-13&59768.47641\\	
59&59909.76523&2022-11-26&59936.18870\\	
60&59936.40397&2022-12-23&59962.08395\\	
	\enddata
\end{deluxetable}

This paper harnesses data from the \textit{Transiting Exoplanet Survey Satellite} (TESS, \citealp{Ricker2015journal}), launched by NASA in 2018 and managed by MIT. NASA's Ames Research Center's Science Processing Operations Center (SPOC, \citealp{Jenkins2016}) is responsible for processing TESS's calibrated data into light curves, providing both Simple Aperture Photometry (SAP) and Pre-Search Data-Conditioned Simple Aperture Photometry (PDCSAP) light curves (for details, see \citealp{2010SPIE.7740E..1UT} and \citealp{2012PASP..124..963K}). Eventually, these data are transmitted to the Mikulski Archive for Space Telescopes (MAST)\footnote{https://mast.stsci.edu/}. Our research has already made significant progress in studying CVs utilizing TESS data (e.g., \citealp{2023arXiv230905891S}; \citealp{Sun2023arXiv}; \citealp{Sun2023MNRAS}; \citealp{2023arXiv230911033S}; \citealp{Sun2024ApJ}).

The photometric observations of HS 2325+8205 were conducted by the TESS mission, encompassing sectors 18, 19, 25, 26, 53, 59, and 60, respectively (refer to Tab. \ref{LOG} and Fig. \ref{light} for details). Our recent research revealed the presence of QPOs during the peak of the long outbursts of HS 2325+8205, utilizing data from sectors 18, 19, 25, and 26 \citep{Sun2023MNRAS}. Subsequently, fresh data from TESS was released, and the current study primarily focuses on analyzing the data obtained from sectors 53, 59, and 60, particularly exploring the impact of the emerging NSHs and and the accretion disk precession on the system.

\begin{figure}
	\centerline{\includegraphics[width=\columnwidth]{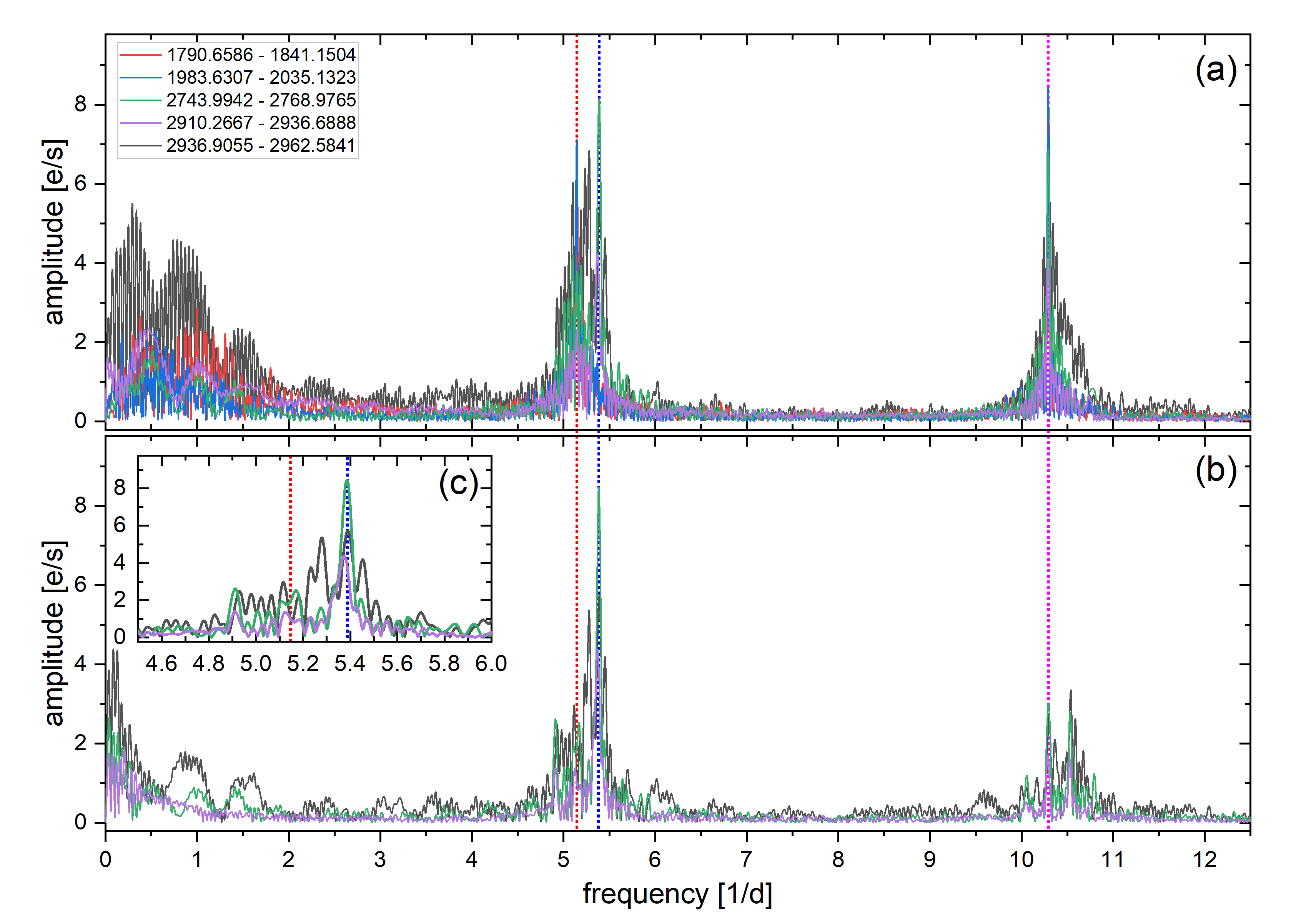}}
	\caption{Frequency spectrograms of HS 2325+8205. (a): Results of the frequency analysis of the light curve after removing the outburst trend by dividing it into five separate parts (see Tab. \ref{data}), the time range in the legend is in units of BJD - 2457000; (b): Results of the analysis of the curves after removing the outburst trend and eclipse for sectors 53, 59 and 60, respectively; (c): Local zoomed-in view of plate (b); vertical red, blue and magenta dashed lines correspond sequentially to the orbital frequency, the NSH frequency and the twice harmonic of the orbital frequency (all averaged over different sectors).\label{fft}}
\end{figure}

\begin{deluxetable*}{lllllllll}
	\tablecaption{Frequency analysis results of Period04 for different curves of HS 2325+8205.\label{data}}
	\tablewidth{0.01pt}
	\tabletypesize{\scriptsize}
	\tablehead{	
		\colhead{Start and End Times$ ^{\rm a} $}&\colhead{Sectors}&\colhead{Types$ ^{\rm b} $}&\colhead{ Frequency }&\colhead{errors }&\colhead{Period}&\colhead{errors}&\colhead{Amplitude$ ^{\rm f} $}&\colhead{errors$ ^{\rm f} $}\\
		\colhead{[d]}&\colhead{}&\colhead{ }&\colhead{[1/d]}&\colhead{[1/d] }&\colhead{[d]}&\colhead{[d]}&\colhead{[e/s]}&\colhead{[e/s]}
	}
	\startdata
	1790.6586 - 1841.1504&18,19&orb&5.1475 &1.63E-04&0.194270 &6.13E-06&6.629 &0.099 
	\\&&2*orb&10.2923 &1.31E-04&0.097160 &1.23E-06&8.253 &0.099 
	\\1983.6307 - 2035.1323&25,26&orb&5.1447 &1.30E-04&0.194376 &4.92E-06&7.111 &0.086  
	\\&&2*orb&10.2910 &1.10E-04&0.097172 &1.04E-06&8.421 &0.086 
	\\2743.9942 - 2768.9765&53&orb&5.1472 &5.79E-04&0.194281 &2.19E-05&4.043 &0.106  
	\\&&nsh&5.3881 &2.82E-04&0.185594 &9.72E-06&8.298 &0.106  
	\\&&2*orb&10.2929 &3.40E-04&0.097155 &3.21E-06&6.892 &0.106  
	\\2910.2667 - 2936.6888&59&orb&5.1437 &7.74E-04&0.194412 &2.93E-05&2.366 &0.088 
	\\&&nsh&5.3729 &4.30E-04&0.186121 &1.49E-05&4.264 &0.088 
	\\&&2*orb&10.2901 &4.45E-04&0.097181 &4.21E-06&4.113 &0.088 
	\\2936.9055 - 2962.5841&60&orb&5.1476 &7.65E-04&0.194265 &2.89E-05&5.671 &0.202 
	\\&&nsh&5.3967 &7.89E-04&0.185299 &2.71E-05&5.500 &0.202 
	\\&&2*orb&10.2943 &5.42E-04&0.097141 &5.12E-06&8.003 &0.202 
	\\2744.3160 - 2962.0519&53,59,60&fre&0.2536 &5.81E-04&3.942596 &9.03E-03&0.067 &0.015 
	\\2744.3160 - 2768.4769&53,&A1&0.2150 &5.05E-03&4.651595 &1.09E-01&2.350 &0.520 
	\\2910.6340 - 2962.0519&59,60&A2&0.2381 &3.11E-03&4.200622 &5.49E-02&1.416 &0.411 
	\\2748.0497 - 2959.0980&53,59,60&O-C&0.2419 &2.65E-04&4.134623 &4.54E-03&0.0010 &0.0001 
	\\2748.0497 - 2959.0980&53,59,60&depth1&0.2421 &2.60E-04&4.131036 &4.43E-03&2.618 &0.260 
	\\2748.0497 - 2959.0980&53,59,60&depth2&0.4842 &4.87E-04&2.065178 &2.08E-03&1.413 &0.263 
	\\2961.2563 - 2962.5730&60&QPO$ ^{\rm c} $&39.477 &2.41E-02 &2188.638 &1.34E+00 &7.054 &0.407  \\
	1251.8247 - 1302.2254&K2(c18)&SDSS J0812$ ^{\rm d} $&0.3214 &3.03E-03 &3.111484 &2.93E-02&0.092 &0.026  \\
	2174.7567 - 2227.0064&32,33&TV Col$ ^{\rm e} $&0.2555 &1.93E-03 &3.913435 &2.96E-02&0.122 &0.022 \\
	\enddata
\begin{threeparttable}
	\begin{tablenotes}[para,flushleft]
		\item [a] The units are BJD - 2457000.\\
	 \item [b] The meanings of the different symbols are as follows: orb = orbital frequency, nsh = NSH frequency, 2*orb = second harmonic of the orbital frequency, fre = frequency of NSH frequency change, A1 = peak frequency of change in NSH amplitude in sector 53, A2 = frequency of NSH amplitude change in sectors 59 and 60, O-C = signal of the O-C curve change, depth1 = peak frequency of eclipse depth, depth2 = peak frequency of the residual from the first analysis (details in Section \ref{subsec:depth}), and QPO = peak frequency of the QPO at the top of the outburst in sector 60.\\
	\item [c] The unit of the QPO's period is the second, and the other period units are days.\\
		\item [d] Periodic analysis of the NSH frequency of SDSS J0812 \citep{Sun2023arXiv} from the 18th observational campaign of K2.\\
		\item [e] Periodic analysis of the NSH frequency of TV Col \citep{Sun2024ApJ}.\\
		\item [f] The symbols fre, SDSS J0812 and TV Col have an amplitude unit of 1/d, O-C has an amplitude unit of d, and all the remaining amplitudes have a unit of e/s.
	\end{tablenotes}
\end{threeparttable}
\end{deluxetable*}


\section{Analysis and Results} \label{sec:Analysis}

\subsection{The emergence of NSHs} \label{subsec:nsh}

To analyze the photometric data of HS 2325+8205 obtained by TESS, we employed the locally weighted regression (LOWESS) method \citep{cleveland1979robust} to remove the outburst trend, as illustrated in Figure \ref{light}. The resulting residuals were segmented into five parts for frequency analysis using Period04 software (details in \citealp{Period04}). Sectors 18 and 19 are treated as one part, sectors 25 and 26 as another, and the remaining 53, 59 and 60 as one part each (see Tab. \ref{data} for the span of each part).
This analysis revealed dominant orbital periods, NSHs, and twice the orbital harmonic signals. Notably, NSHs were detected specifically in sectors 53, 56, and 60, as shown in Figures \ref{fft}b and c. Further analysis excluding eclipse phases emphasized the prominence of NSHs. The average period of NSHs was determined to be 0.185671(17) d, averaged over the three sectors (see Fig. \ref{fft}c and Tab. \ref{data}). The orbital period averaged over all five sectors was calculated as 0.194321(18) d. This marks the first detection of NSHs in HS 2325+8205.
The excess was calculated as $\epsilon = -0.0445(2)$, where $\rm \epsilon = (P_{nsh} - P_{orb})/P_{orb}$. 
No SORs were identified in the light curve. The theoretical precession signal was calculated to be 4.171(17) d (0.240(1) d$^{-1}$, $\rm 1/P_{prec} = 1/P_{nsh} - 1/P_{orb}$).

\subsection{Parameter Extraction for NSHs} \label{subsec:Extraction}

More recent studies have highlighted the complexity and variability of NSHs, where a single period is insufficient to describe its behavior. We observed periodic variations in NSH amplitude, O-C, and eclipse depth in NLs and IPs, and we aimed to verify this phenomenon in DN HS 2325+8205, where no SORs were found. If periodic variations similar to the tilted disk precession cycle were found, it would be possible to demonstrate that the tilted disk has not disappeared. Therefore, we employed a segmented fitting approach similar to \cite{Sun2023arXiv, Sun2024ApJ} to explore the evolution of NSHs.

\begin{figure}
	\centerline{\includegraphics[width=\columnwidth]{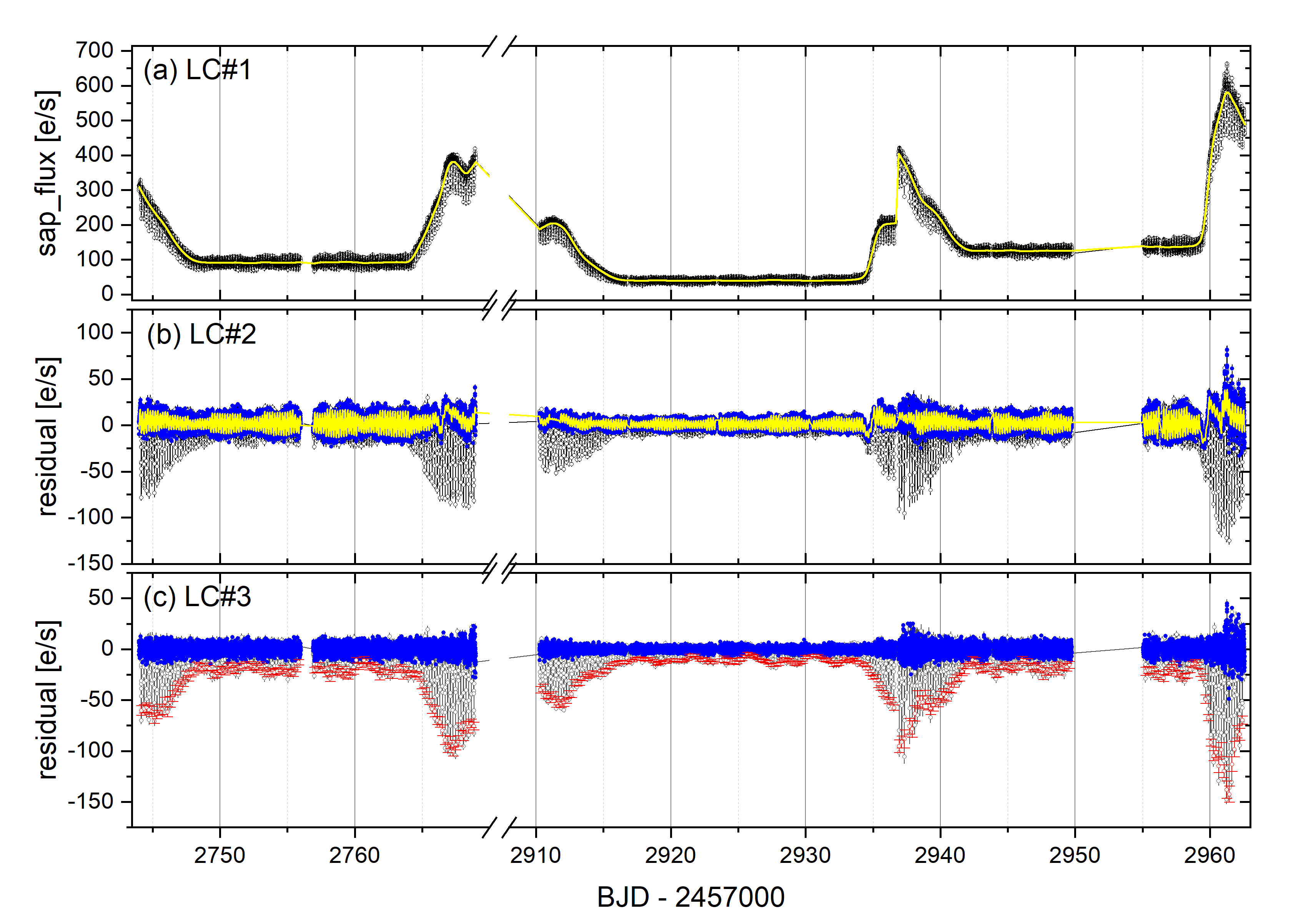}}
	\caption{Segmented fitting process. (a) sap\_flux light curves (LC\#1) for sectors 53, 59, and 60, the yellow solid line is the LOWESS fit; (b) residuals (LC\#2) of the LOWESS fit to LC\#1, the blue points are the out-of-eclipse curves, and the yellow solid line is the segmental fit to the out-of-eclipse curves; (c) residuals (LC\#3) obtained by subtracting the yellow solid line from all light curves in LC\#2, the blue points are the out-of-eclipse curves and the red points are the eclipse minima.\label{de_nsh}}
\end{figure}

NSHs were detected solely in sectors 53, 56, and 60, with the corresponding SAP\_flux light curve labeled LC\#1 (see Fig. \ref{de_nsh}a). The light curve after removing the outburst trend was denoted as LC\#2 (see Fig. \ref{de_nsh}b). The out-of-eclipse portion of LC\#2 was divided into 91 segments, each spanning approximately 4 times the orbital period ($\sim$ 0.78 days; accounting for potential data gaps of varying durations, see Tab. \ref{tab:A1}). Following the methods of \cite{Sun2023arXiv, Sun2024ApJ}, we initially attempted a segmented linear superposition sinusoidal model to fit the out-of-eclipse curve of LC\#2 (depicted by blue circles in Fig. \ref{de_nsh}b), but encountered poor fitting results during outburst peaks. Due to the persistent outburst presence in LC\#1, which hindered accurate LOWESS trend removal per segment, we opted for a least-squares quadratic superimposed sinusoidal model without weighting to mitigate outburst effects:

\begin{equation} \label{eq:sine}
	\rm Flux(BJD) = Z + B \cdot BJD + C \cdot BJD^2 + A \cdot \sin(2\pi \cdot (freq \cdot BJD + p))
\end{equation}
Here, $\rm Z$, $\rm B$, $\rm C$, $\rm A$, $\rm freq$, and $\rm p$ represent the fitted intercept, linear coefficients, quadratic coefficients, amplitude, frequency, and phase, respectively. The theoretical NSH curve derived from the segmental fit was subtracted from the LC\#1 light curve to obtain LC\#3, which represents the light curve after removing the NSH trend. The final fit results are depicted by the yellow solid line in Figure \ref{de_nsh}b and summarized in Table \ref{tab:A1}, providing crucial insights into NSH amplitude and frequency.

\subsection{Frequency Variation} \label{subsec:Frequency}

\begin{figure}
	\centerline{\includegraphics[width=\columnwidth]{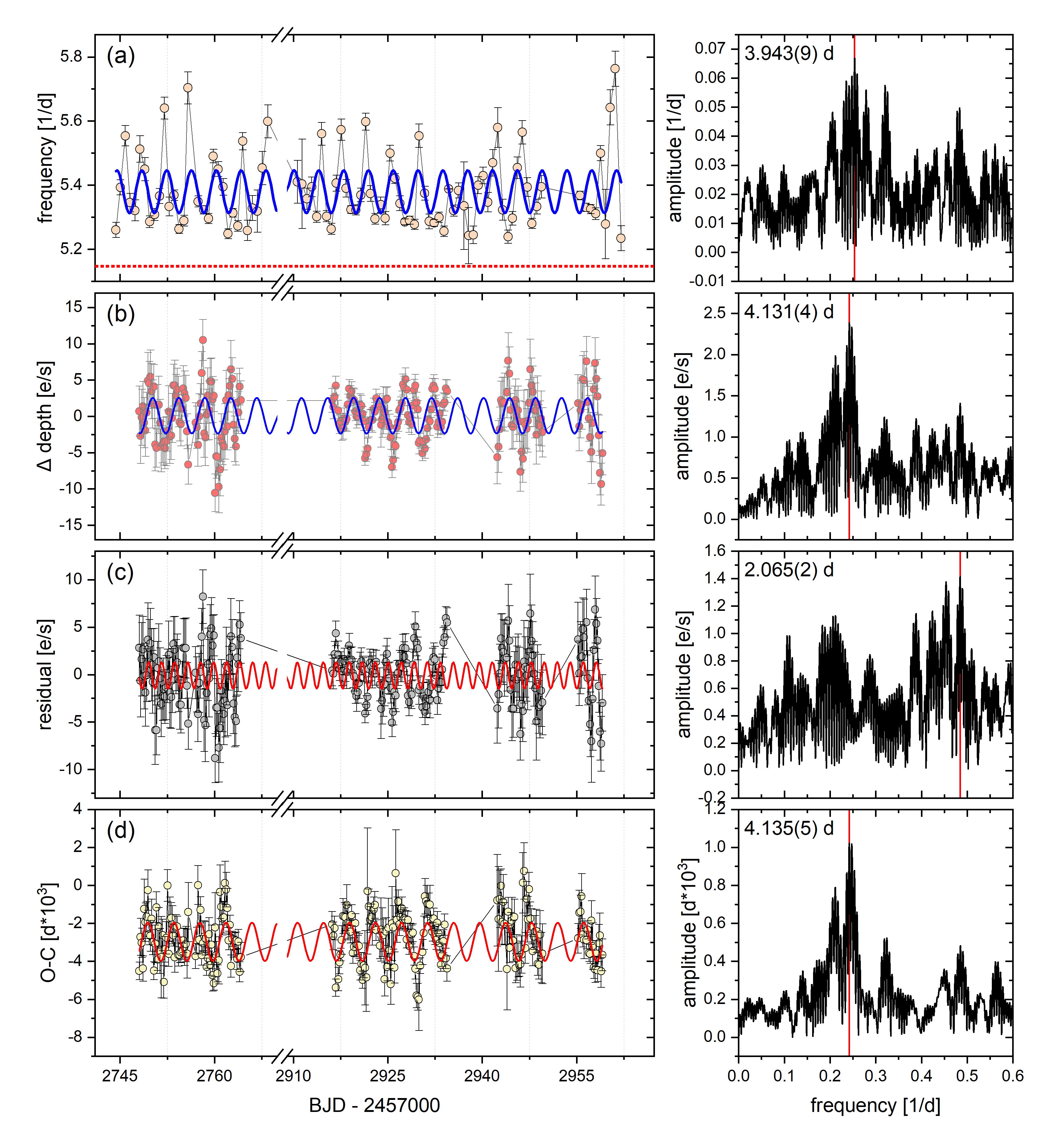}}
	\caption{Frequency analysis of different curves of HS 2325+8205. Right panel: frequency spectra of the different curves, the peak frequencies have been marked using red vertical lines; Left panel: the scattered dots are the different curves, and the solid line is the sinusoidal fit corresponding to the peak frequencies (see Tab. \ref{data} for parameters); (a), (b), (c), and (d) are in turn the NSH frequencies, the eclipse depths, and the residuals from the fitting of the (b) panels, O-C between the observed and calculated eclipse minimum epochs. The red horizontal dashed line in (a) is the orbital frequency. \label{4fft}}
\end{figure}

The average value of BJD for each segment served as the x-axis for plotting, yielding curves depicting NSH frequency and amplitude variations over time. The frequencies of NSHs were consistently higher than the orbital frequency, exhibiting periodic instabilities and variations (see Fig. \ref{4fft}a). Analysis conducted using Period04 revealed that NSH frequencies exhibit periodic variations with a period of 3.943(9) d (see Fig. \ref{4fft}a and Tab. \ref{data}). The sinusoidal fits corresponding to the peak NSH frequency are represented by blue solid lines in Fig. \ref{4fft}a (Period04 software integrates frequency analysis with sinusoidal fitting, details in \citealp{Period04}).

To visualize these periodic variations more intuitively, we selected the light curve from sector 59 as an example and zoomed in (see Fig. \ref{example}). Using BJD 2459760.6121 as the zero-phase point (which was standardized for all subsequent folding), the NSH frequency changes were folded over phases 0 to 2, revealing the periodic changes in NSH frequency (see Fig. \ref{flod}a).
This study marks the first observation of periodic variations in NSH frequencies resembling the theoretical cycle of tilted disk precession, providing crucial observational evidence for understanding the origin of NSHs.

\subsection{Amplitude Variation} \label{subsec:Amplitude}

We have observed periodic variations in the NSH amplitude similar to the tilted disk precession period in NLs (e.g., SDSS J0812, \citealp{Sun2023arXiv}) and IPs (e.g., TV Col, \citealp{Sun2024ApJ}). To verify the generality of this phenomenon, we conducted a similar analysis of NSH amplitude in the DN HS 2325+8205. The NSH amplitudes obtained from segmented fits exhibited both positive and negative values, with absolute values used for analysis. 
The analysis of NSH amplitude revealed instability characterized by various types of variations:

(a) Quadratic variation in NSH amplitude in sector 53 (see Fig. \ref{CWT}a);

(b) Long-term increase in NSH amplitude observed in sectors 59 and 60 (see Fig. \ref{CWT}a);

(c) Periodic variations in amplitude across sectors 53, 59 and 60 (see Fig. \ref{CWT}b);

(d) Amplitude of periodic variations in NSH amplitude increasing in sectors 59 and 60 over time (see Figs. \ref{CWT}a and b).

To determine the period of periodic amplitude variation, we performed parabolic fit to NSH amplitude data in sector 53 and linear fit in sectors 59 and 60 (blue and red solid lines in Fig. \ref{CWT}a), respectively, to mitigate long-term variations. The residual data from these fits were separately analyzed for frequency (see Figs. \ref{CWT}e and f). The periods of amplitude variation were found to be P$\rm _{|A|, s53} $ = 4.652(109) d and P$\rm _{|A|, s59, s60} $ = 4.201(55) d (see Tab. \ref{data}). The period of sector 53 is slightly longer than the theoretical precession period, while sectors 59 and 60 exhibit variations closer to the theoretical precession period, a bias that may be related to the small number of data points in sector 53.

Consistent with the approach of \cite{Sun2024ApJ}, we employed Continuous Wavelet Transform (CWT; \citealp{CWT1989}) to validate changes in NSH amplitude. The segmented fitting of the out-of-eclipse curve of LC\#2 provided the parabolic and sinusoidal components, yielding LC\#4 after removing the quadratic component (dotted line plot in Fig. \ref{CWT}c) and retaining only the sinusoidal component (red solid line in Fig. \ref{CWT}c). CWT applied to LC\#4 confirmed long-term and periodic variations in NSH amplitude. The colors of the stripes in the two-dimensional power spectrum of CWT indicate NSH intensity. Focusing on sector 59 (2910.2667 - 2936.6888, BJD - 2457000), combined with Figs. \ref{CWT}a, b, c, significant enhancement in NSH intensity and clear periodic variation in NSH amplitude were observed, consistent with segmental fitting results.

This study represents the first observation of periodic variations in NSH amplitude similar to the tilted disk precession period in DNe, underscoring the general applicability of this phenomenon.

\subsection{O-C} \label{subsec:O-C}

The Observation minus Calculation (O-C) method of eclipse minima provides an effective means to study the evolution of binary star systems. In CVs, primary stars, including white dwarfs, accretion disks, and hot spots, are eclipsed by secondary stars when the orbital inclination exceeds a critical threshold. The luminosity of white dwarfs remains stable, with variations primarily attributed to the accretion disk. Eclipse minima represent the centers of brightness during these eclipses. Recent studies have indicated that eclipse minima in CVs exhibit variations similar to timescales associated with accretion disk precession. Therefore, O-C analysis of eclipse minima serves as an ideal method to investigate accretion disk variations in eclipsed CVs.

\begin{figure*}
	\centerline{\includegraphics[width=1.2\columnwidth]{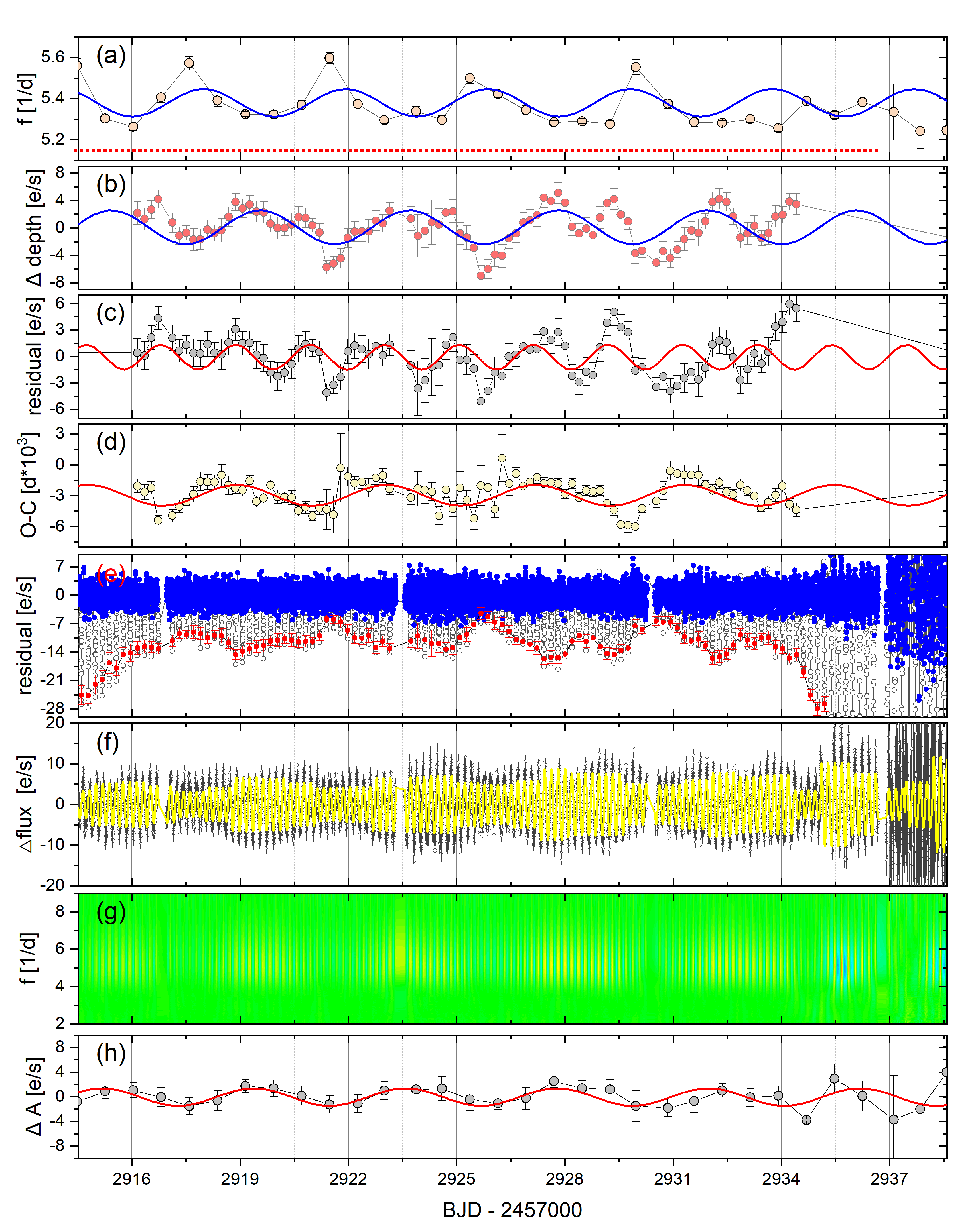}}
	\caption{Details of different curves and their contrasts. The data was selected from sector 59 in order to clearly show the details of the changes and to compare the different curves. (a): the frequency variation of NSHs is consistent with Fig. \ref{4fft}a; (b): the variation curve of eclipse depth is in accordance with Fig. \ref{4fft}b; (c): in agreement with Fig. \ref{4fft}c; (d): the O-C variation curve is in agreement with Fig. \ref{4fft}d; (e): the LC\#3 light curve; (f): the LC\#4 light curve; (g): CWT 2D power spectra of LC\#4 is consistent with Fig. \ref{CWT}d; (h): the variation curve of the amplitude of NSHs is in agreement with Fig \ref{CWT}b.
		\label{example}}
\end{figure*}

Previously, we utilized the LOWESS and segmented fitting methods to remove outburst and negative superhump (NSH) trends, resulting in LC\#3 (see Fig. \ref{de_nsh}c). Consistent with \cite{Sun2023arXiv}, we computed 356 eclipse minima for HS 2325+8205 using a linear superposition Gaussian fit (see Tab. \ref{tab:A2}). An O-C analysis of these quiescent minima was carried out on the basis of ephemeris derived from \cite{Sun2023MNRAS}:

\begin{equation}\label{eq:ephemeris}
	\rm	Min.I = BJD 2458793.6762(6) + 0.19433475(6) \times E  
\end{equation}
where E represents the cycle number. The O-C analysis revealed significant periodic variations, with frequency analysis determining periods and amplitudes of 4.135(5) days and 87.3(89) seconds, respectively (see spectrum in Fig. \ref{4fft} and sinusoidal fit to peak frequency in Tab. \ref{data}), consistent with theoretical accretion disk precession periods. 
Further observation showed that the O-C variation is non-sinusoidal, with a rise time approximately half that of the decline (see Figs. \ref{4fft}c and \ref{example}d). The folded O-C curve is depicted in Figure. \ref{flod}e, where instead of a sinusoidal fit, we binned the curve at 0.03 phase intervals, further revealing that the O-C curve rises faster than it declines.

\subsection{Eclipse Depth} \label{subsec:depth}

Using a method consistent with \cite{Sun2023arXiv,Sun2024ApJ}, we computed the variation in eclipse depth for HS 2325+8205, utilizing the -0.5 to -0.1 and 0.1 to 0.5 phases of each eclipse from the LC\#3 light curves as out-of-eclipse data. To exclude outburst effects, we focused on data during quiescence and performed linear fits to the eclipse depths for sectors 53, 59, and 66, yielding the final relative eclipse depth variation curves (see Fig. \ref{4fft} b). Frequency analysis of these eclipse depths revealed a periodic variation with a peak frequency of 0.2421(3) d$^{-1}$ (P$_{1}$ = 4.131(4) d). The sinusoidal fit corresponding to this peak frequency is depicted as blue solid lines in Figures. \ref{4fft}b and \ref{example}b. Further frequency analysis of residuals from the sinusoidal fit identified a periodic variation with a period of P$_{2}$ = 2.065(2) d, approximately half that of P$_{1}$ (see Figs. \ref{4fft}c and \ref{example}c).

\begin{figure}
\centerline{\includegraphics[width=\columnwidth]{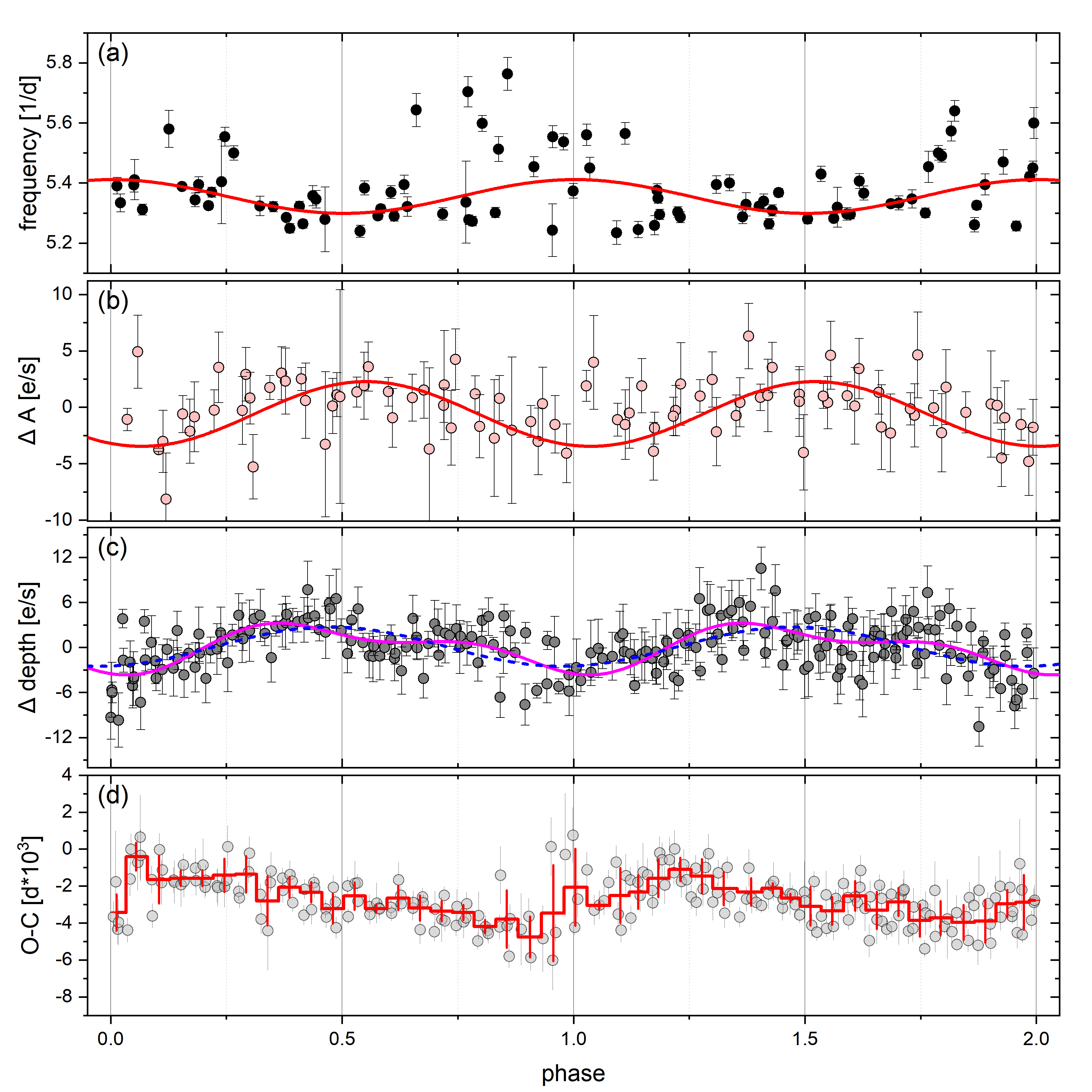}}
\caption{Different folding curves for HS 2325+8205. The zero-phase point is BJD2459760.6121, and the folding period corresponds to the frequency analysis results of the different curves (see Tab. \ref{data}); (a): folded curves for the frequency variation of the NSHs, with the red solid line being a sinusoidal fit; (b): folded curves for the amplitude variation of the NSHs, with the red solid line being a sinusoidal fit; (c): folded curves for the variation of the eclipse depth, with blue dashed and magenta solid lines being a single-sinusoidal and a double-sinusoidal fit, respectively; (d): folded curves for O-C, the red curves are the result of binning at 0.03 phase intervals.\label{flod}}
\end{figure}

Additionally, following \cite{Sun2024ApJ}, we folded the eclipse depth curve and applied both double-sine and single-sine fits. The results confirmed the presence of a bi-periodic variation in the eclipse depth (see Fig. \ref{flod}c and d). Upon detailed comparison of the variations in LC\#3 and the eclipse depth curves, the clear bi-peak structure is evident in the minima (see Figs. \ref{example}b and e).

This discovery represents the first observation of a bi-periodic variation in eclipse depth in non-magnetic CVs, providing crucial observational evidence for understanding tilted disk precession in CVs.

\subsection{QPOs} \label{subsec:qpo}

\begin{figure}
\centerline{\includegraphics[width=\columnwidth]{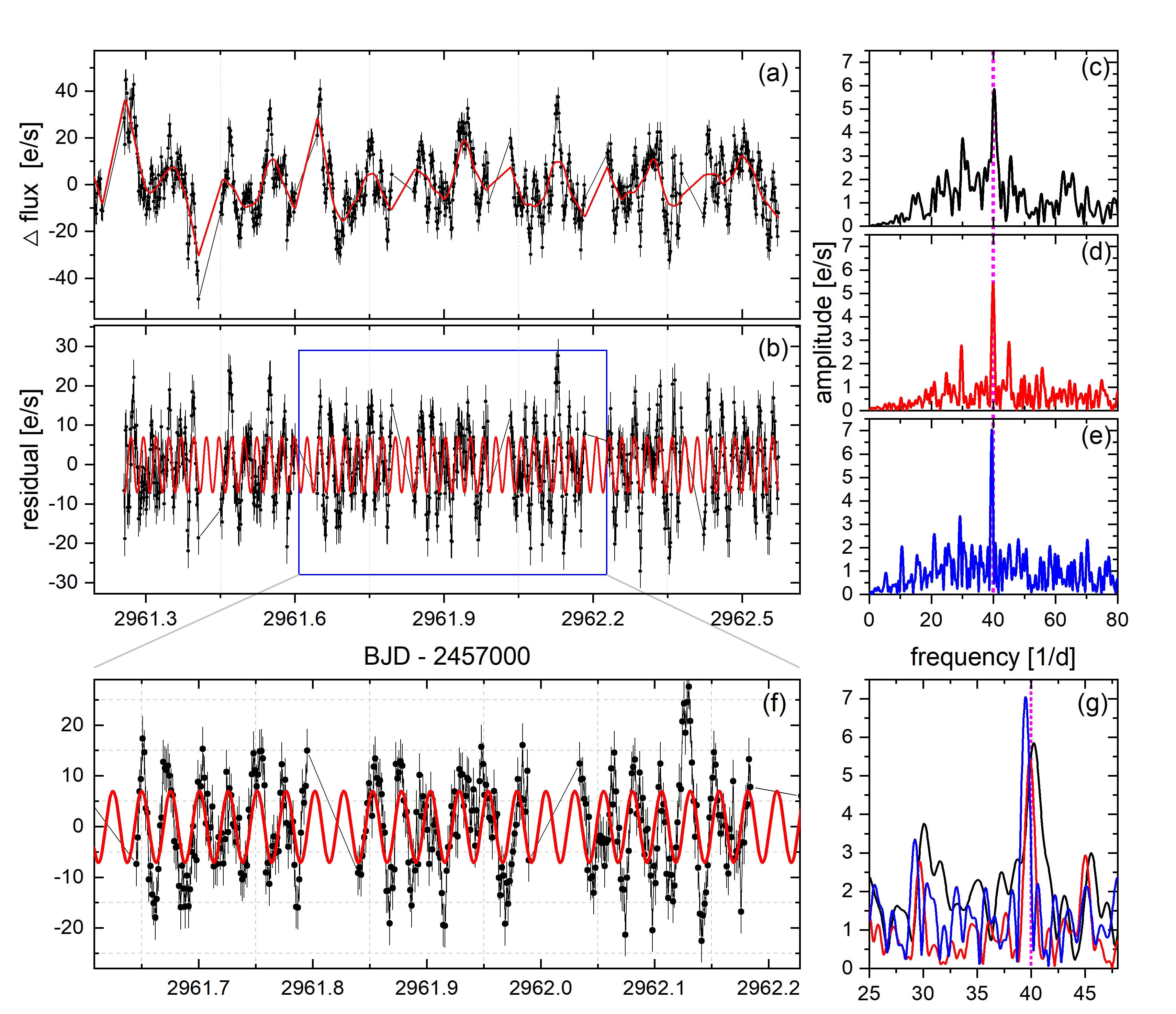}}
\caption{QPOs for HS 2325+8205 in sector 60. (a): outburst top of sector 60 corresponding to LC\#3 curve, red solid line is the LOWESS fit; (b): residuals of the LOWESS fit, red solid line is the sinusoidal fit corresponding to the peak frequency; (c), (d) and (e) are comparisons of the spectrograms of the three QPOs light curves corresponding to the times in order 1827.1297 - 1827.9867, 1994.2584 - 1995.6320 and 2961.2563 - 2962.5730, and the magenta dashed line corresponds to a frequency of 40 d$ ^{-1}$; (f): enlarged view of the light curves in the blue-coloured rectangular box in panel (b); (g): combined view of panels (c), (d) and (e), with the curve colours corresponding to the individual panels in turn.\label{QPO}}
\end{figure}

In recent work, QPOs were identified with frequencies of 40.263(5) d$^{-1}$ ($\sim$ 2145.89 s, 1827.1297 - 1827.9867) and 39.871(3) d$^{-1}$ ($\sim$ 2166.99 s, 1994.2584 - 1995.6320) during the peak of two long outbursts in HS 2325+8205 \citep{Sun2023MNRAS}. In this study, QPOs were similarly detected at the peak of a long outburst in sector 60. Following the methodology of \cite{Sun2023MNRAS}, the top curve was selected to remove the orbital trend using a LOWESS fit (see Fig. \ref{QPO}a), and frequency analysis of the residuals identified the period of the QPOs as 39.477(24) d$^{-1}$ ($\sim$ 2188.62 s, 2961.2563 - 2962.5730), generally consistent with previous observations (see Figs. \ref{QPO}). Spectrograms of the three datasets were plotted to compare the three QPOs, showing no significant changes over the maximum time span of QPO detection (see Figs. \ref{QPO}c, d, e and g), approximately 1135 days ($ \sim $ 3 years). This stable period of QPOs in HS 2325+8205 suggest that QPOs may have originated from physical changes on timescales longer than at least $ \sim $ 3 years, providing  important observational evidence for studying the origin and characteristics of QPOs.

In addition, a comparison of the three periodograms reveals not only the presence of QPOs but also two distinct signals at frequencies of approximately 30 d$ ^{-1} $ and 45 d$ ^{-1} $. The periodogram from the interval 1994.2584 to 1995.6320 is the most significant among them (see Fig. \ref{QPO}d). Using Figure \ref{QPO}d as an example, these signals are identified as f$ _1 $ = 29.784(56) d$ ^{-1} $ and f$ _2 $ = 45.176(70) d$ ^{-1} $. These signals might arise from the combined frequencies of the orbital frequencies and QPOs, f$ _1 $ $ \approx $ (39.871(3) - 2$\rm *f_{orb} $) d$^{-1}$ and f$ _2 $ $ \approx $ (39.871(3) + $\rm  f_{orb}$) d$^{-1}$. The combined frequencies are still interesting to study, and the reasons for why  - 2$\rm *f_{orb}$ and + $\rm f_{orb}$ are observed rather than other combinations need to be explored further.

\begin{figure}
	\centerline{\includegraphics[width=\columnwidth]{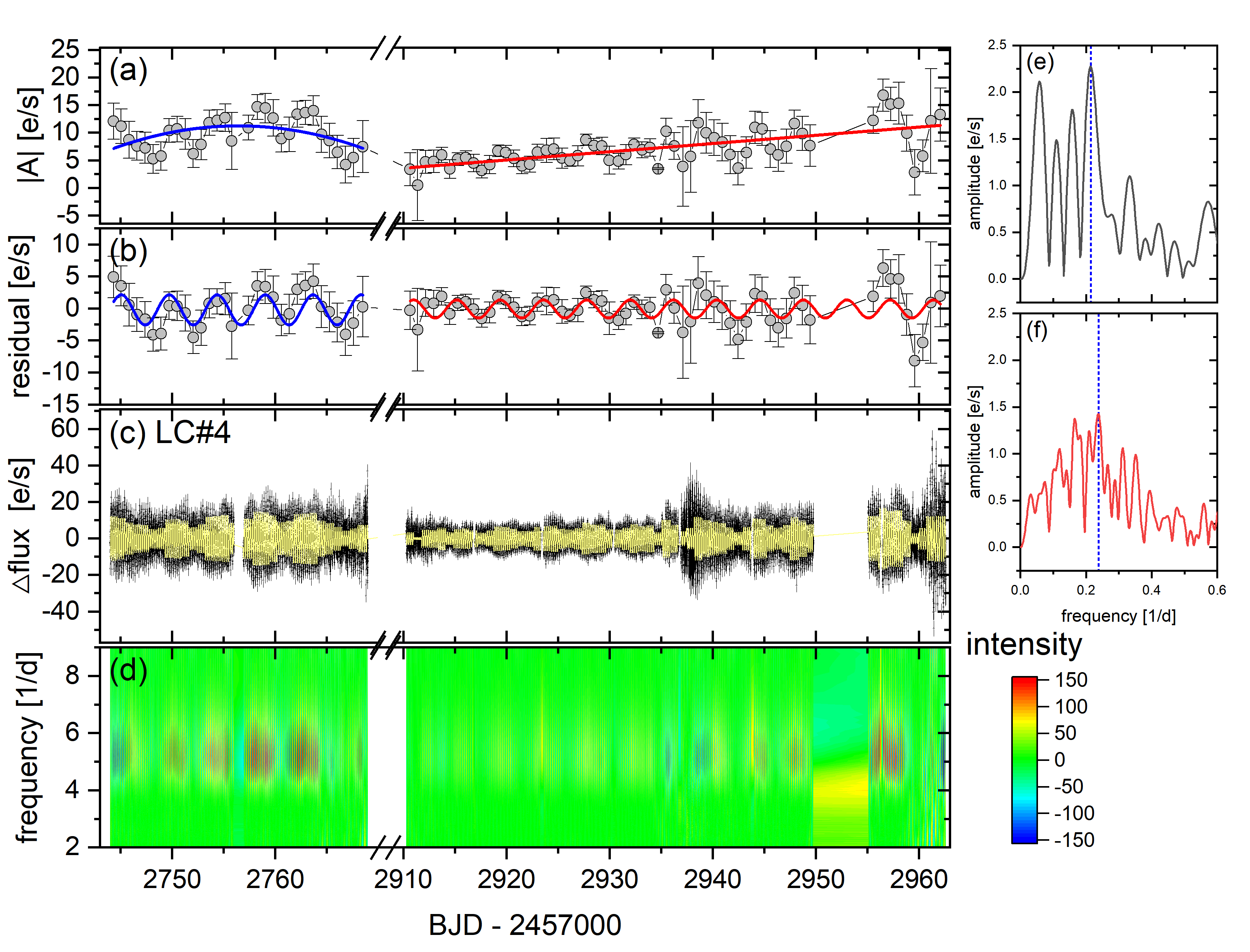}}
	\caption{Analysis of the amplitudes of NSHs. (a): NSHs amplitude variation curves, blue and red lines are parabolic and linear fits, respectively;
		(b): residuals of the fit in layout (a), blue and red lines corresponding to sinusoidal fits to the peak frequencies of the frequency analysis of the NSH amplitudes in sector 53, sectors 59 and 60, respectively;
		(c): out-of-eclipse curves of LC\#2 with the parabolic component removed from the segmental fit, the yellow solid line is the sinusoidal part of the segmental fit;
		(d): CWT 2D power spectrum of LC\#4.
	(e) and (f): results of the frequency analyses of the NSHs amplitudes in sector 53, sectors 59 and 60, corresponding to the black and red solid lines, respectively, with the peak frequencies marked by vertical dashed lines.\label{CWT}}
\end{figure}

\section{DISCUSSION} \label{sec:DISCUSSION}

\subsection{Correlation of Different Windows} \label{sec:Correlation}

We observed that the NSH frequency, NSH amplitude, eclipse minima, and eclipse depth exhibit cyclic variations similar to the theoretical period of tilted disk precession. These findings suggest that although there are no SORs, the tilted disk precession is still present, and the question of why the SORs are not significant is still to be answered. 

\begin{figure}
	\centerline{\includegraphics[width=\columnwidth]{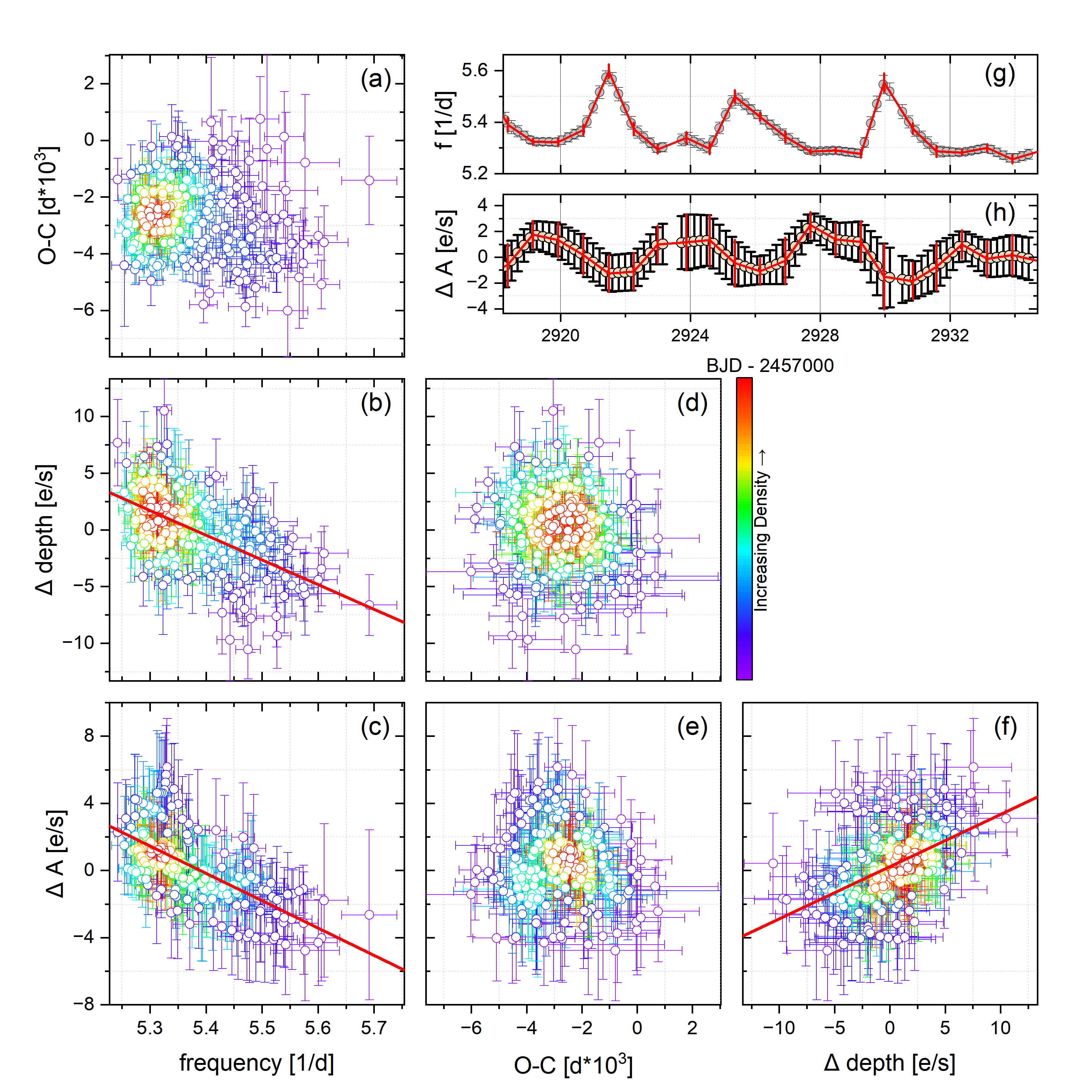}}
	\caption{Correlation of different curves. (a), (b), (c), (d), (e), and (f) in order of NSH frequency vs. O-C, NSH frequency vs. eclipse depth, NSH frequency vs. NSH amplitude, O-C vs. eclipse depth, O-C vs. NSH amplitude, O-C vs. NSH amplitude, and eclipse depth vs. NSH amplitude. The colours in the figure indicate the density (amount of data per unit area), with redder colours indicating more data points per unit area. (g) and (h): examples and results of linear interpolation of NSHs frequency and amplitude, the dotted line in the figure is the result of interpolation and the red line plot is the raw data. \label{Correlation}}
\end{figure}

In this Section, we explore their interrelationships and intrinsic significance.
In Section \ref{subsec:Extraction}, NSH information was derived via segmented fitting, yielding 91 data points. To enhance our analysis, O-C and eclipse depth evaluations were extended to encompass 227 points during the quiescent state. Direct comparison between NSH information and O-C or eclipse depth was not feasible.
To proceed, we initially focused on quiescent NSH data and employed linear interpolation using eclipse minima as the base BJD to expand the data points for NSH frequency and NSH amplitude to 227 points. This method generated frequency and amplitude change curves for NSHs corresponding to each eclipse minima. Comparison between interpolated results and original curves is depicted in Figures. \ref{Correlation}g and h, demonstrating consistent trends.

Figure \ref{Correlation} illustrates relationships among the four variation curves. A significant correlation is evident between NSH frequency, NSH amplitude, and eclipse depth, whereas O-C shows no significant relationship with the other variables. Linear fitting results reveal:

(a) NSH amplitude decreases as NSH frequency increases (see Fig. \ref{Correlation}c), confirming the opposite evolutionary trend between the two variables observed in Figures. \ref{Correlation}g and h.

(b) Similarly, eclipse depth decreases with increasing NSH frequency (see Fig. \ref{Correlation}b), displaying notable changes as shown in Figures. \ref{example}a and b, with sinusoidal curves demonstrating opposite evolutionary trends.

(c) Conversely, eclipse depth increases with NSH amplitude (see Fig. \ref{Correlation}f and Fig. \ref{example}b and h), providing further evidence for (a) and (b) (where both NSH amplitude and eclipse depth decrease with frequency, suggesting isotropic evolution of NSH amplitude and eclipse depth).

These findings offer crucial observational evidence for subsequent discussions on their origin and for investigating tilted disk precession and NSHs.

\subsection{Complex Variations in NSH Amplitude}

First, we utilize NSHs amplitude variations as a starting point to explore various new phenomena. Before delving into these variations, let us revisit the origins of NSHs. Numerous studies support the notion that NSHs stem from periodic fluctuations in the energy released from streams impacting a retrogradely precessing titled disk (e.g., \citealp{1985A&A...143..313B}; \citealp{harvey1995superhumps}; \citealp{Patterson1997PASP}; \citealp{patterson1999permanent}; \citealp{Wood2007ApJ...661.1042W}; \citealp{wood2009sph}; \citealp{Montgomery2009MNRAS}; \citealp{Montgomery2012}; \citealp{2013MNRAS.435..707A}; \citealp{thomas2015emergence}; \citealp{KimuraKIC94066522020PASJ}; \citealp{BruchIII}; \citealp{Sun2024ApJ}). Because the tilted disk is slowly reverse precession, when a fast binaries turn a circle, the streams from the secondary star will have the opportunity to enter the farthest point of the inner disk twice (two faces of the tilted disk), producing two humps with periods slightly less than half the orbital period; due to the large optical thickness of the disk, the observer will only be able to see a single hump during one orbit.
The release of TESS data has provided a valuable opportunity to study the origin of NSHs, leading to the discovery of numerous new phenomena, which pose both challenges and opportunities for models of reverse precession of tilted disks.

In Section \ref{subsec:Amplitude}, we identify four variations characterizing the NSHs amplitude (see Fig. \ref{CWT}; (a) parabolic variation, (b) long-term increase, (c) periodic variation, (d) long-term increase in the amplitude of periodic variation of NSH amplitude). To explain these phenomena, we enumerate the main sources of NSH-related optical variations in CVs:

(i) Thermal instability on the accretion disk leading to a luminosity rise (outburst; e.g., \citealp{1996PASP..108...39O}; \citealp{lasota2001disc}; \citealp{2020AdSpR..66.1004H}).

(ii) Changes in the radius of the accretion disk (e.g., \citealp{1983A&A...128...29V}; \citealp{1989PASJ...41.1005O}; \citealp{warner1995cat}), resulting in shifts in the impact point of the stream and consequent alterations in the released kinetic energy.

(iii) Variations in the matter transfer rate from the secondary star (e.g., \citealp{Smak2013}; \citealp{2017A&A...602A.102H}; \citealp{2020AdSpR..66.1004H}).

(iv) Changes in the tilt angle of the accretion disk (e.g., \citealp{Montgomery2009MNRAS}; \citealp{Osaki2013b}; \citealp{KimuraKIC94066522020PASJ}).

(v) Retrograde precession of the tilted disk, origin of SORs.

(vi) Variation in the hot spot position as the material stream traverses the disk, origin of NSHs.

(vii) Variations caused by the combination of binary motion phase changes and tilted disk precession phase changes due to unchanged observational positioning.

Recent investigations have shown in DNe AH Her, ASAS J1420, TZ Per, and V392 Hya that NSH amplitudes decrease during rising outbursts and increase during declining outbursts \citep{2023arXiv230905891S, 2023arXiv230911033S}, interpreted as arising from changes in the accretion disk radius (Variable source: ii). In this study, we observe a parabolic variation in NSH amplitude for HS 2325+8205 in sector 53. Sector 53's light curve displays two incomplete outbursts at its start and end (see Fig. \ref{de_nsh}a, 2743.9942 - 2768.9765), with the middle period in quiescence. Additionally, NSHs in sector 53 exhibit their highest amplitudes during quiescence, decreasing during rising outbursts and increasing during declining outbursts, mirroring the behavior observed in AH Her. 
Consequently, we suggest that the parabolic variation in NSH amplitude in Sector 53 stems from changes in the accretion disk radius (Variable source: ii). During outburst rise, the disk radius expands, reducing the distance from the Lagrangian L1 point to the disk contact point, thereby diminishing the energy release from the material. Conversely, during outburst decline, the disk contracts. However, the absence of significant parabolic variation in sectors 59 and 60 is perplexing and might be obscured by longer-term, more pronounced variations.

Periodic variations in NSH amplitude have been observed in NL SDSS J0812 \citep{Sun2023arXiv} and IP TV Col \citep{Sun2024ApJ}, and this phenomenon is noted for the first time in DNe. This periodic variation provides strong evidence for the presence of a tilted disk with reverse precession in NSHs. Consistent with interpretations of SDSS J0812 and TV Col, the origin of periodic amplitude variation may be akin to the origin of SORs, stemming from the periodic variation in the apparent hot spot area due to tilted disk precession (Variable source: vii).

Additionally, we have newly identified that the amplitude of cyclic variations in NSH amplitude increases with a long-term increase in NSH amplitude. The amplitudes of NSHs in sectors 59 and 60 exhibit a prolonged increase beyond the outburst timescale, potentially associated with changes in the tilt angle of the disk (Variable source: iv) and the rate of matter transfer (Variable source: iii). The relationship between the orbital and NSHs and the tilted disk precession frequency can be expressed as \citep{1995MNRAS.274..987P,1998MNRAS.299L..32L,Osaki2013b,Smak2013,KimuraKIC94066522020PASJ,2023MNRAS.520.3355S}:
\begin{equation} \label{eq:cos}
	\nu_{\rm nsh} - \nu_{\rm orb} =  \nu_{\rm prec} = -\frac{3}{8\pi} \frac{GM_{2}}{a^{3}} \frac{\int \Sigma r^{3}dr}{\int \Sigma\Omega r^{3}dr}  \rm cos\theta
\end{equation}
$ 	\nu_{\rm nsh}   $: Orbital frequency;
$ \nu_{\rm orb}   $: NSH frequency;
$ 	\nu_{\rm prec}  $ : Tilted disk precession frequency;
$ 	G  $: Gravitational constant;
$ 	M_{2}  $ : Mass of the secondary;
$ 	a   $: Binary separation;
$ 	\Sigma   $: Surface density of the disk;
$ 	\Omega  $ : Keplerian angular velocity of the disk matter;
$ 	r  $: Radial distance from the accretion disk to the white dwarf;
$ 	\theta   $: Tilt angle.
Assuming other parameters are constant, an increase in the tilt angle leads to an increase in $\nu_{\rm nsh}$, but our analysis of NSH frequencies reveals no significant long-term increasing trend except for periodic variations. Hence, as noted by \cite{Smak2013}, this may be related to a long-term increase in the matter transfer rate (Variable source: iii). If there is a sustained increase in the material transfer rate, this could also explain the long-term increase in the amplitude of periodic NSH amplitude variations, which are larger for the same orbital phase and tilted disk precession phase.

\subsection{Not only IPs but also DNe} \label{subsec:NSH and SOR}

Periodic variations in eclipse depth related to tilted disk precession have been observed in NL DW UMa \citep{Boyd2017} and NL \citep{2021MNRAS.503.4050I}. In our recent studies, we have detected periodic variations in eclipse depth in both IP TV Col \citealp{Sun2024ApJ} and NL SDSS J0812 \citep{Sun2023arXiv}. TV Col exhibits bi-periodic variations with periods $\rm P_{1} = 3.905(11)$ d and $\rm P_{2} = 1.953(4)$ d, where $\rm P_{1} \approx P_{\text{prec}} \approx 2 \times P_{2}$. We interpreted $\rm P_{1}$ as evidence of tilted disk precession, attributing it to periodic changes in the apparent area of the tilted disk. The occurrence of $\rm P_{2}$ exclusively in IPs, not NLs, led us to propose that this bi-periodic variation might be linked to the precession of two accretion curtains along with the tilted disk. However, this hypothesis has evolved with the findings presented in this paper.

	In our recent study, we present the first evidence of bi-periodic variations in eclipse depth in the DN HS 2325+8205 ($\rm P_{1} = 4.131(4)$ d, $\rm P_{2} = 2.065(2)$ d, $\rm P_{\text{prec}} = 4.171(17)$ d), resembling those observed in TV Col. Interestingly, this bi-periodic variation in eclipse depth is not exclusive to IPs. The amplitude of $\rm P_{1}$ in HS 2325+8205 is approximately twice that of $\rm P_{2}$ (see Table \ref{data}), suggesting that $\rm P_{1}$ originates from a brighter source compared to $\rm P_{2}$. $\rm P_{1}$ closely matches the theoretical precession period of the tilted disk, implying periodic changes in the brightness of the eclipsing center due to tilted disk precession, further supporting the evidence of tilted disk precession. Additionally, we found a correlation where eclipse depth increases with the amplitude of NSHs (see Figs. \ref{Correlation}f, \ref{example}b, and \ref{example}h), consistent with our interpretation of cyclic variations in NSH amplitude and eclipse depth with $\rm P_{1}$. As the apparent area of the hot spot increases, so does the eclipse depth.

Surprising and confusing is the change in period for $\rm P_{2}$. However, when this phenomenon was first noticed in the IP TV Col, we suggested that it was due to the accretion curtain precessing with the tilted disk \citep{Sun2024ApJ}. Yet, the same variation was found in the non-magnetic DN HS 2325+8205, which showed no relation to the accretion curtain.
$\rm P_{2}$ is approximately half of the theoretical period of tilted disk precession, suggesting a clear association with tilted disk precession. Simply put, during one complete precession cycle of the tilted disk ($\rm P_{prec}$), the tilted disk reveals an additional bright source appearing twice, or two sources appearing once each, in addition to the brightness of the disk itself. A similar change in tilted disk precession is the fact that the stream does have the opportunity to enter the furthest point twice on two faces of the tilted disk. This implies that the additional brightness source is likely still due to the motion of the hot spot across the disk face. 

Additionally, the observation that this bi-periodic variation is evident in the eclipse depth but not in NSH amplitude, NSH frequency, or O-C highlights a distinct difference in the origins of these variations. To fully understand this, it is crucial to consider or rule out various possible factors, including the type of binary system, orbital inclination, fluctuations in the accretion disk's transparency, and the distribution of material, among others.
Detailed studies are scarce, and additional observational evidence is necessary to validate our hypotheses. We also caution that the bi-periodic variation may alternatively manifest as a bi-peak variation in eclipse depth.

\subsection{New evidence for the origin of NSHs} 

\begin{figure}
	\centerline{\includegraphics[width=\columnwidth]{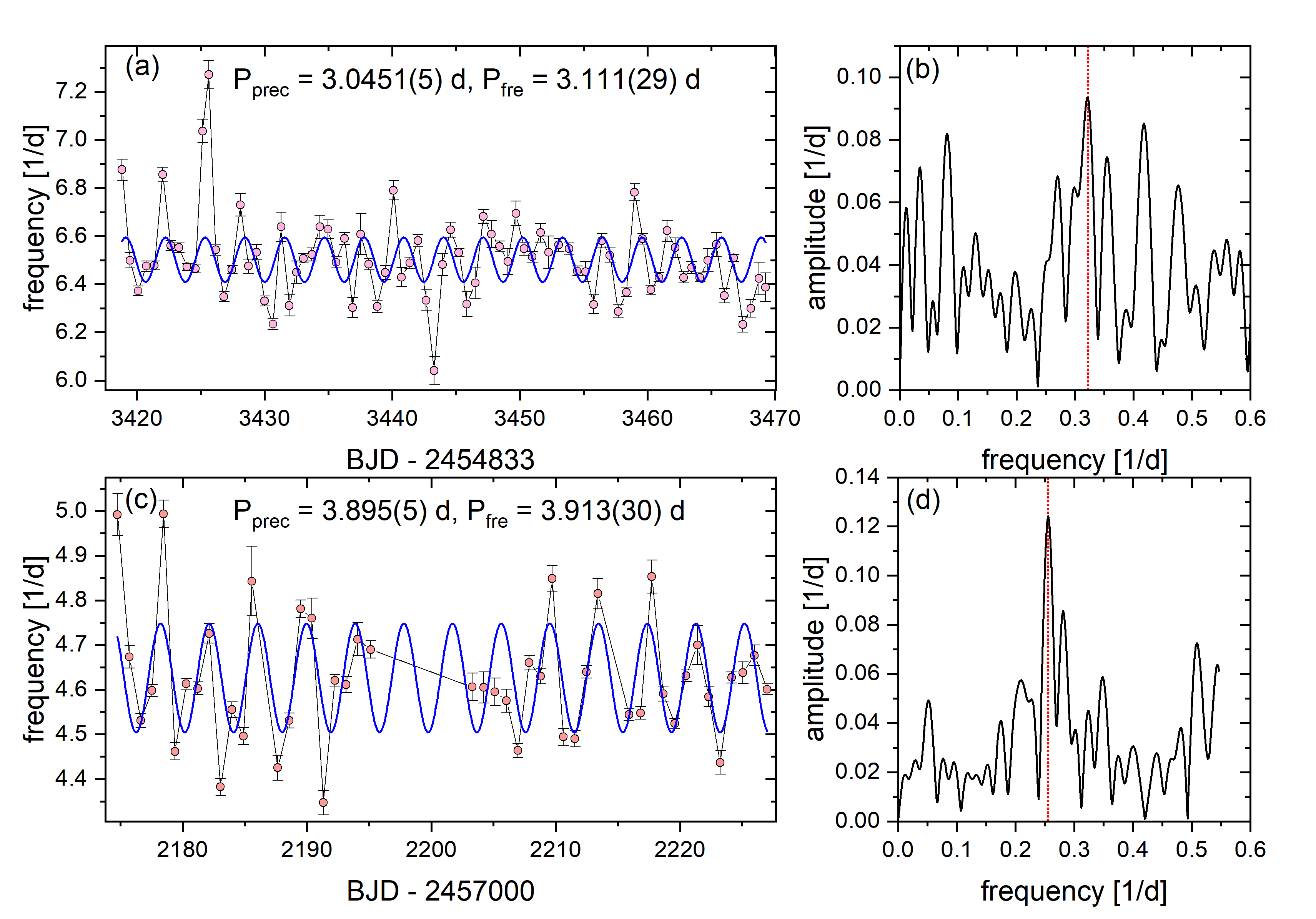}}
	\caption{NSH frequency variation curves and frequency analysis of SDSS J0812 and TV Col. The data presented are sourced from \cite{Sun2023arXiv} for SDSS J0812 and \cite{Sun2024ApJ} for TV Col. In each subplot, the vertical solid line indicates the peak frequency, and the blue solid line represents the sinusoidal fit corresponding to this peak frequency. (a) and (b) correspond to the NSH frequency variations and corresponding frequency analyses for SDSS J0812, while (c) and (d) correspond to the NSH frequency variations and corresponding frequency analyses for TV Col.\label{frequency}}
\end{figure}

In Section \ref{subsec:Frequency}, we identify a periodic variation in the frequency of NSHs in DN HS 2325+8205 ($\rm P_{fre}$ = 3.943(9) d), which closely matches the theoretical period of tilted disk precession, observed here for the first time. To verify this phenomenon we try to seek validation in SDSS J0812 and TV Col.
Frequency analyses were conducted using segmented fit parameters reported in \cite{Sun2023arXiv} and \cite{Sun2024ApJ}. The results reveal periodic variations in NSH frequency with periods of $\rm P_{fre}$ = 3.111(29) d ($\rm P_{prec}$ = 3.0451(5) d; \citealp{Sun2023arXiv}) and $\rm P_{fre}$ = 3.913(30) d ($\rm P_{prec}$ = 3.895(5) d; \citealp{Sun2024ApJ}) in SDSS J0812 and TV Col, respectively (see Fig. \ref{frequency}). This establishes that periodic variations in NSH frequency, approximately matching the period of tilted disk precession, are common across NLs, IPs, and DNe.

The NSH frequency variation can be described by Eq. \ref{eq:cos}, where the orbital and precession frequencies are pivotal. Additionally, we observe a periodic variation in the O-C diagram of eclipse minima in HS 2325+8205 ($\rm P_{O-C}$ = 4.135(5) d and $\rm A_{O-C}$ = 87.3(89) s, see Section \ref{subsec:O-C}), characterized by a faster rise than fall. This could potentially contribute to the observed frequency variations of NSHs, given that O-C variations stem from changes in the orbital period.
Unfortunately, such short-term periodic O-C variations cannot be attributed to dominant long-term magnetic activity cycles \citep{Applegate1992mechanism}, the light travel-time effect caused by a third body (\citealp{irwin1952determination}; \citealp{borkovits1996invisible}), or slight amplitude fluctuations of the accretion disk \citep{schaefer2021fast}. Instead, they likely represent periodic oscillations of the eclipse center due to accretion disk precession \citep{Miguel2016accretion, Sun2023arXiv, Sun2024ApJ}.

If NSH frequency variations are indeed linked to changes in the period of tilted disk precession, periodic fluctuations in the precession period would be expected. However, the SOR cycle remains stable in both SDSS J0812 and TV Col. Moreover, the possibility of alternating acceleration and deceleration within a single precession cycle of the tilted disk could potentially produce asymmetric signals in the SORs (constant SOR period but with asymmetric non-sinusoidal changes in shape). Yet, no such asymmetry was detected in the nearly sinusoidal variations of the SORs observed in SDSS J0812 and TV Col. Therefore, we propose that NSH frequency variations are not solely a product of the precession cycle.

Crucial observations include $\rm P_{fre} \approx P_{prec}$, the NSH amplitude decreasing with increasing NSH frequency, and the inverse correlation observed between eclipse depth and NSH frequency (see Section \ref{sec:Correlation} and Fig. \ref{Correlation}). This suggests that if NSH frequency variation is not directly related to the tilt disk precession period, it may correlate with the precession phase. We suggest that such variations may arise from an interaction between orbital phase and tilted disk precession phase, similar to the variations observed in NSH amplitude and eclipse depth. However, the exact mechanism remains puzzling and requires further study.

\section{CONCLUSIONS} \label{sec:CONCLUSIONS}

In this paper, we present an in-depth study of the tilted disk precession and the origin of the NSHs, based on TESS photometry, using an analysis similar to that of \cite{Sun2023arXiv} and \cite{Sun2024ApJ}, with a sample of long-period eclipsing DN HS 2325+8205. We utilize the eclipsing minima O-C, the eclipsing depth, the NSH frequency, NSH amplitude and the correlation between them as the windows. The main results are as follows:

(1) For the first time, we found NSHs with period $0.185671(17)$ d ($\epsilon = -0.0445(2)$) in HS 2325+8205, but no SORs. Discontinuities in NSHs and the absence of co-occurrence of SORs and NSHs in HS 2325+8205 provide observational evidence for studying the formation and cessation of tilted disks and NSHs. Additionally, we rediscovered QPOs with a period of $\sim 2188.62$ s ($39.477(24)$ d$^{-1}$) in the new observational data, with the period remaining stable for over 3 years. This provides crucial observational evidence for understanding the origin of QPOs.

(2) In HS 2325+8205 a similar periodic variation in the NSH frequency ($\rm P_{fre}$ = 3.943(9) d) was found as in the tilted disk precession cycle. Subsequently, we verified this in NL SDSS J0812 ($\rm P_{fre}$ = 3.111(29) d) and IP TV Col ($\rm P_{fre}$ = 3.913(30) d) and again found similar periodic variations, suggesting that this phenomenon is general. In addition, we found that both NSH amplitude and eclipse depth decreased with increasing NSH frequency. Therefore we suggest that the periodic variations in NSH frequency may have a similar origin to the periodic variations in NSH amplitude and eclipse depth from the interaction of the orbital phase with the tilted disk precession phase.

(3) The NSH amplitude exhibits complex variations, including (a) parabolic variation in sector 53, (b) long-term increase in sectors 59 and 60, (c) periodic variation across all sectors, and (d) long-term increase in the amplitude of periodic variation of NSH amplitude in sectors 59 and 60. We suggest that (a) may resemble DNe AH Her, ASAS J1420, TZ Per, and V392 Hya, attributable to variations in the accretion disk radius, (b) may be associated with variations in matter transfer rate, (c) mirrors the origin of SORs from cyclic variations in hot spot apparent area, and (d) may result from combined variations in matter transfer rate and hot spot apparent area.

(4) Similarly, we found periodic variations in the eclipsing minima O-C of HS 2325+8205, with periods and amplitudes $\rm P_{O-C} = 4.135(5)$ d and $\rm A_{O-C} = 87.3(89)$ s, respectively. These variations may be linked to periodic changes in the eclipsing center in response to tilted disk precession. Additionally, we observed that the rate of rise of O-C is about half that of the rate of fall, indicating asymmetric brightness distribution of the tilted disk.

(5) Analogous to IP TV Col, we identified bi-periodic variations in eclipse depth with periods $\rm P_{1} = 4.131(4)$ d and $\rm P_{2} = 2.065(2)$ d ($\rm P_{1} \approx P_{\text{prec}} \approx 2 \times P_{2}$) in DN HS 2325+8205. This is the first observation of such cycles in DNe, suggesting that the variation with period $\rm P_{2}$ may not be related to accretion curtain. We suggest that the periodic variation with period $\rm P_{2}$ may also arise from material streams impacting the furthest points of the two sides.

We conducted a detailed study on tilted disk precession and NSHs in DN HS 2325+8205 across multiple observation windows. Our results indicate the presence of periodic variations approximately matching the precession period in O-C, eclipse depth, NSH frequency, and NSH amplitude. These findings strongly support the existence of tilted disk precession, although no SORs were detected, and the question of why the SORs are not significant urgently needs to be addressed.
The most intricate variations observed are in NSH frequency, NSH amplitude, and eclipse depth, and there is a correlation between them. This paper provides important observational evidence on these phenomena, which is key evidence for the study the origin of the NSHs.

\section*{Acknowledgements}

This work was supported by National Key R\&D Program of China (grant No. 2022YFE0116800), the National Natural Science Foundation of China (Nos. 11933008). In the compilation of this paper, we have relied heavily on the invaluable data gathered by the TESS survey, and we express our heartfelt appreciation to all the diligent personnel involved in this esteemed mission. Additionally, we recognize with gratitude the Mikulski Archive for Space Telescopes (MAST) at the Space Telescope Science Institute for hosting and facilitating access to all the data presented herein.
All of the TESS data used in this paper can be found through the MAST: \dataset[doi:10.17909/6xrj-6k75]{https://doi.org/10.17909/6xrj-6k75}

\bibliography{sample631}{}
\bibliographystyle{aasjournal}

\section{Appendix: supplementary material on-line}\label{sec:Appendix}
The appendix contains Tables \ref{tab:A1} and \ref{tab:A2}. Table \ref{tab:A1} shows the segmented parabolic superimposed sinusoidal fitting parameters, calculated as described in Section \ref{subsec:Extraction} of the main paper. Table \ref{tab:A2} shows the quiescent eclipsing parameters for HS 2325+8205, including eclipsing minima, depth, E and O-C analyses.

\renewcommand\thetable{A1}
\setlength{\tabcolsep}{1pt}
\begin{longtable}{ccccccccccccccccccc}

	\caption{Parameters obtained from fitting the NSHs using a parabolic superimposed sine fit.}\label{tab:A1}\\

	
	\toprule
	Start$^a$&End$^a$&during&Z&err&B&errors&C&errors&A&errors&freq&errors&p&errors&$\rm \nu$$^b$&$\rm \chi_{red}^2$$^c$\\

[BJD]&[BJD]&[d]&[e/s]&[e/s]&[$\rm e / (s \cdot d) $]&[$\rm e / (s \cdot d) $]&[$\rm e / (s \cdot d^2) $]&[$\rm e / (s \cdot d^2) $]&[e/s]&[e/s]&[1/d]&[1/d]&[rad] &[rad] &&\\
	\midrule
	\endfirsthead
	\multicolumn{17}{c}
	{{\bfseries{Table \thetable\ }continued from previous page}} \\
	\toprule
	Start$^a$&End$^a$&during&Z&err&B&errors&C&errors&A&errors&freq&errors&p&errors&$\rm \nu$$^b$&$\rm \chi_{red}^2$$^c$\\

[BJD]&[BJD]&[d]&[e/s]&[e/s]&[$\rm e / (s \cdot d) $]&[$\rm e / (s \cdot d) $]&[$\rm e / (s \cdot d^2) $]&[$\rm e / (s \cdot d^2) $]&[e/s]&[e/s]&[1/d]&[1/d]&[rad] &[rad] &&\\
\midrule
	\endhead
	\midrule
	\multicolumn{17}{c}
	{{ }} \\
	\endfoot
	\endlastfoot
2743.9942&2744.6484&0.65&35.43&3.93&-1.94E+05&2.16E+04&2.67E+08&2.96E+07&12.105&3.235&5.261&0.024&350.46&64.92&368&21.29 
\\2744.6498&2745.4262&0.78&2.88&0.98&-1.58E+04&5.40E+03&2.17E+07&7.41E+06&11.185&3.117&5.393&0.023&-13.41&63.79&433&23.92 
\\2745.4275&2746.2026&0.78&-12.73&0.77&6.99E+04&4.21E+03&-9.60E+07&5.78E+06&8.710&3.334&5.554&0.032&-454.19&87.24&429&26.19 
\\2746.2039&2746.9803&0.78&-3.70&1.89&2.03E+04&1.04E+04&-2.79E+07&1.43E+07&7.616&2.603&5.346&0.029&117.15&79.85&430&18.51 
\\2746.9817&2747.7581&0.78&5.16&0.90&-2.83E+04&4.93E+03&3.89E+07&6.78E+06&7.286&2.789&5.321&0.032&184.07&89.08&432&16.62 
\\2747.7595&2748.5345&0.78&3.30&0.94&-1.81E+04&5.15E+03&2.49E+07&7.07E+06&5.258&2.598&5.513&0.042&-340.81&114.47&429&17.87 
\\2748.5359&2749.3123&0.78&3.40&1.33&-1.87E+04&7.34E+03&2.57E+07&1.01E+07&5.779&2.554&5.450&0.036&-168.24&99.56&430&13.41 
\\2749.3137&2750.0901&0.78&2.16&1.41&-1.19E+04&7.75E+03&1.63E+07&1.07E+07&10.423&2.167&5.286&0.018&281.54&49.86&426&15.14 
\\2750.0915&2750.8665&0.78&0.08&1.42&-4.56E+02&7.81E+03&6.28E+05&1.07E+07&10.677&2.322&5.308&0.019&219.90&51.43&428&15.34 
\\2750.8679&2751.6443&0.78&9.32&0.86&-5.13E+04&5.06E+03&7.06E+07&3.04E+06&9.784&2.797&5.366&0.024&60.75&65.19&431&16.44 
\\2751.6457&2752.4207&0.78&4.58&0.90&-2.52E+04&5.23E+03&3.47E+07&3.41E+06&6.215&2.531&5.640&0.034&-692.07&94.10&431&17.06 
\\2752.4221&2753.1985&0.78&13.97&0.43&-7.69E+04&2.37E+03&1.06E+08&3.26E+06&7.865&2.758&5.334&0.029&150.08&80.97&430&15.99 
\\2753.1999&2753.9763&0.78&7.61&0.86&-4.19E+04&3.15E+03&5.77E+07&1.21E+07&11.828&2.287&5.369&0.016&54.06&44.78&431&14.76 
\\2753.9776&2754.7527&0.78&-5.78&1.27&3.19E+04&7.02E+03&-4.39E+07&9.66E+06&12.233&1.951&5.263&0.014&343.29&38.32&430&12.31 
\\2754.7540&2755.5304&0.78&3.57&1.27&-1.97E+04&7.00E+03&2.71E+07&9.65E+06&12.719&2.512&5.289&0.017&272.98&46.57&430&14.71 
\\2755.5318&2755.9930&0.46&95.65&0.68&-5.27E+05&5.29E+03&7.26E+08&7.29E+06&8.496&5.167&5.704&0.050&-868.25&138.76&265&17.73 
\\2756.9124&2757.8624&0.95&7.64&0.96&-4.21E+04&5.29E+03&5.81E+07&7.29E+06&10.921&2.236&5.349&0.017&108.17&46.94&524&21.33 
\\2757.8638&2758.6402&0.78&-1.83&0.85&1.01E+04&4.66E+03&-1.39E+07&6.43E+06&14.675&2.231&5.323&0.013&181.60&35.84&433&15.97 
\\2758.6416&2759.4166&0.78&-3.32&0.68&1.83E+04&3.78E+03&-2.53E+07&5.21E+06&14.474&2.665&5.296&0.016&254.99&43.47&429&18.96 
\\2759.4180&2760.1944&0.78&-1.67&0.72&9.19E+03&4.00E+03&-1.27E+07&5.51E+06&12.693&3.322&5.490&0.021&-281.49&58.67&430&22.68 
\\2760.1958&2760.9722&0.78&11.45&1.08&-6.32E+04&5.96E+03&8.72E+07&8.23E+06&8.969&2.471&5.450&0.023&-169.72&63.74&433&17.39 
\\2760.9736&2761.7486&0.78&-0.98&1.41&5.39E+03&7.79E+03&-7.44E+06&1.07E+07&9.717&3.105&5.395&0.026&-17.28&71.93&429&19.51 
\\2761.7500&2762.5264&0.78&-9.63&1.22&5.32E+04&6.72E+03&-7.34E+07&9.28E+06&13.346&2.338&5.249&0.015&386.13&42.07&430&16.80 
\\2762.5278&2763.3028&0.78&7.80&1.46&-4.31E+04&8.07E+03&5.95E+07&1.11E+07&13.671&2.167&5.314&0.014&206.96&37.37&424&14.43 
\\2763.3042&2764.0806&0.78&-5.09&0.74&2.82E+04&4.07E+03&-3.89E+07&5.63E+06&14.025&2.676&5.273&0.016&319.25&45.33&429&18.47 
\\2764.0820&2764.8584&0.78&-16.19&2.34&8.95E+04&1.29E+04&-1.24E+08&1.79E+07&9.739&3.260&5.537&0.027&-411.05&75.27&431&23.17 
\\2764.8598&2765.6348&0.78&-3.95&0.51&2.18E+04&2.85E+03&-3.02E+07&3.94E+06&8.566&3.131&5.259&0.032&360.21&87.47&430&28.91 
\\2765.6362&2766.4126&0.78&-86.57&0.43&4.79E+05&2.37E+03&-6.62E+08&3.28E+06&6.508&3.029&5.329&0.039&165.12&108.17&430&22.89 
\\2766.4140&2767.1904&0.78&-127.30&0.49&7.04E+05&2.72E+03&-9.75E+08&3.77E+06&4.239&3.334&5.319&0.066&191.87&182.82&432&26.90 
\\2767.1918&2767.9668&0.78&-29.87&0.45&1.65E+05&2.47E+03&-2.29E+08&3.42E+06&5.507&3.459&5.454&0.052&-180.49&144.06&429&30.67 
\\2767.9682&2768.9765&1.01&44.98&14.80&-2.49E+05&8.19E+04&3.45E+08&1.13E+08&-7.484&4.711&5.599&0.052&-584.12&142.82&566&74.14 
\\2910.2667&2910.9890&0.72&33.58&4.33&-1.95E+05&2.52E+04&2.84E+08&3.67E+07&3.397&3.014&5.411&0.067&-65.36&195.02&392&16.10 
\\2910.9903&2911.7667&0.78&-7.28&0.21&4.24E+04&1.19E+03&-6.17E+07&1.74E+06&-0.495&6.432&5.404&0.139&1.51&1.75&430&10.84 
\\2911.7681&2912.5445&0.78&-14.18&0.60&8.26E+04&3.50E+03&-1.20E+08&5.09E+06&4.749&2.082&5.357&0.033&90.38&96.50&430&9.30 
\\2912.5459&2913.3209&0.78&-9.50&0.27&5.54E+04&1.59E+03&-8.06E+07&2.32E+06&-4.818&2.020&5.395&0.031&-19.64&89.85&432&7.32 
\\2913.3223&2914.0987&0.78&8.59&0.85&-5.01E+04&4.94E+03&7.29E+07&7.20E+06&6.025&1.367&5.302&0.017&252.84&50.87&430&5.00 
\\2914.1001&2914.8765&0.78&-2.03&0.31&1.18E+04&1.82E+03&-1.73E+07&2.65E+06&3.445&1.699&5.560&0.036&-500.76&104.59&430&4.77 
\\2914.8778&2915.6528&0.78&1.02&0.31&-5.93E+03&1.80E+03&8.64E+06&2.62E+06&5.227&1.172&5.303&0.017&248.86&50.19&431&3.73 
\\2915.6542&2916.4306&0.78&-6.30&0.97&3.67E+04&5.67E+03&-5.36E+07&8.27E+06&5.509&1.227&5.264&0.017&363.77&50.64&429&3.64 
\\2916.4320&2917.2084&0.78&2.52&0.60&-1.47E+04&3.51E+03&2.14E+07&5.12E+06&4.522&1.560&5.406&0.025&-52.13&73.39&314&3.58 
\\2917.2098&2917.9848&0.78&10.32&0.22&-6.02E+04&1.28E+03&8.79E+07&1.87E+06&3.174&1.389&5.573&0.033&-538.11&96.08&429&4.18 
\\2917.9862&2918.7626&0.78&10.74&0.31&-6.27E+04&1.79E+03&9.15E+07&2.61E+06&4.222&1.635&5.391&0.028&-6.22&80.50&429&4.24 
\\2918.7640&2919.5403&0.78&-2.05&0.77&1.20E+04&4.51E+03&-1.75E+07&6.59E+06&6.696&1.054&5.324&0.012&187.15&34.97&430&2.91 
\\2919.5417&2920.3167&0.78&4.29&0.27&-2.51E+04&1.56E+03&3.66E+07&2.28E+06&6.405&1.335&5.323&0.016&191.98&46.31&431&3.85 
\\2920.3181&2921.0945&0.78&6.11&0.33&-3.57E+04&1.94E+03&5.21E+07&2.83E+06&5.351&1.529&5.369&0.021&55.26&61.37&430&3.65 
\\2921.0959&2921.8709&0.78&4.56&0.45&-2.67E+04&2.65E+03&3.90E+07&3.87E+06&4.019&1.414&5.598&0.026&-612.74&77.18&430&4.23 
\\2921.8723&2922.6487&0.78&4.34&0.39&-2.54E+04&2.31E+03&3.71E+07&3.37E+06&4.302&1.447&5.374&0.025&42.98&72.17&432&3.27 
\\2922.6501&2923.3195&0.67&1.53&0.26&-8.92E+03&1.53E+03&1.30E+07&2.24E+06&6.514&1.440&5.295&0.017&273.82&49.53&384&3.83 
\\2923.5653&2924.2028&0.64&5.42&0.33&-3.17E+04&1.96E+03&4.63E+07&2.86E+06&6.810&2.166&5.339&0.024&143.76&70.51&357&6.48 
\\2924.2042&2924.9806&0.78&0.53&0.65&-3.08E+03&3.81E+03&4.49E+06&5.58E+06&7.077&1.936&5.297&0.021&266.98&60.83&432&6.57 
\\2924.9820&2925.7584&0.78&-1.71&0.34&1.00E+04&2.00E+03&-1.46E+07&2.93E+06&5.414&1.825&5.500&0.025&-326.80&72.30&430&6.33 
\\2925.7598&2926.5348&0.78&-0.74&0.19&4.34E+03&1.12E+03&-6.36E+06&1.64E+06&4.905&0.990&5.421&0.016&-94.92&45.41&430&3.18 
\\2926.5361&2927.3125&0.78&7.95&0.34&-4.66E+04&1.97E+03&6.81E+07&2.88E+06&5.831&1.804&5.344&0.023&131.26&67.10&431&5.67 
\\2927.3139&2928.0903&0.78&-0.80&0.54&4.69E+03&3.17E+03&-6.87E+06&4.63E+06&8.719&1.018&5.285&0.009&301.63&26.16&430&2.78 
\\2928.0917&2928.8667&0.77&-6.11&0.46&3.58E+04&2.72E+03&-5.24E+07&3.98E+06&7.710&1.268&5.290&0.013&286.94&36.83&431&3.66 
\\2928.8681&2929.6444&0.78&-0.63&0.38&3.68E+03&2.21E+03&-5.40E+06&3.24E+06&7.631&1.589&5.278&0.016&324.65&46.56&431&4.37 
\\2929.6458&2930.2792&0.63&22.78&1.03&-1.33E+05&6.03E+03&1.96E+08&8.84E+06&5.040&2.533&5.554&0.037&-484.99&107.12&358&6.42 
\\2930.4958&2931.1986&0.70&8.97&0.42&-5.26E+04&2.44E+03&7.70E+07&3.58E+06&4.844&1.414&5.375&0.022&38.94&63.30&379&3.07 
\\2931.2000&2931.9764&0.78&3.44&0.63&-2.02E+04&3.67E+03&2.96E+07&5.38E+06&6.057&1.822&5.287&0.023&297.39&66.64&431&6.08 
\\2931.9778&2932.7541&0.78&-1.10&0.58&6.43E+03&3.40E+03&-9.43E+06&4.98E+06&7.908&1.108&5.282&0.011&311.60&31.38&430&3.30 
\\2932.7555&2933.5305&0.77&-8.18&0.92&4.80E+04&5.42E+03&-7.04E+07&7.95E+06&6.892&1.432&5.300&0.016&258.48&46.29&431&4.63 
\\2933.5319&2934.3083&0.78&-10.82&0.34&6.35E+04&1.99E+03&-9.31E+07&2.93E+06&7.320&1.566&5.256&0.016&387.06&48.25&430&4.50 
\\2934.3097&2935.0861&0.78&124.38&0.34&-7.30E+05&1.97E+03&1.07E+09&2.88E+06&3.488&0.150&5.388&0.001&0.00&1.69&430&4.97 
\\2935.0874&2935.8624&0.77&-42.71&1.52&2.51E+05&8.93E+03&-3.68E+08&1.31E+07&10.290&2.372&5.321&0.017&198.06&51.00&432&14.06 
\\2935.8638&2936.6402&0.78&-7.78&0.39&4.57E+04&2.29E+03&-6.71E+07&3.36E+06&7.600&2.449&5.383&0.024&14.70&69.25&430&10.62 
\\2936.6416&2937.4180&0.78&-47.15&1.77&2.77E+05&1.04E+04&-4.07E+08&1.53E+07&3.902&7.193&5.336&0.137&154.25&401.17&306&45.19 
\\2937.4194&2938.1943&0.77&-31.84&0.23&1.87E+05&1.38E+03&-2.75E+08&2.02E+06&5.698&6.495&5.243&0.087&429.60&257.00&343&85.74 
\\2938.1957&2938.9721&0.78&25.31&2.49&-1.49E+05&1.46E+04&2.19E+08&2.15E+07&11.814&4.149&5.245&0.027&421.03&79.49&430&48.14 
\\2938.9735&2939.7499&0.78&-1.99&0.55&1.17E+04&3.24E+03&-1.72E+07&4.76E+06&10.015&3.673&5.400&0.027&-34.41&78.41&430&22.84 
\\2939.7513&2940.5262&0.77&2.51&0.88&-1.48E+04&5.18E+03&2.17E+07&7.61E+06&9.115&3.217&5.430&0.026&-122.63&76.34&431&23.37 
\\2940.5276&2941.3040&0.78&-4.51&6.69&2.65E+04&3.93E+04&-3.90E+07&5.78E+07&8.309&3.383&5.347&0.030&125.27&88.71&430&20.35 
\\2941.3054&2942.0818&0.78&0.74&0.86&-4.37E+03&5.07E+03&6.43E+06&7.46E+06&-6.042&3.457&5.470&0.041&-239.98&120.83&430&19.12 
\\2942.0832&2942.8581&0.77&16.63&2.65&-9.79E+04&1.56E+04&1.44E+08&2.29E+07&-3.602&3.018&5.580&0.062&-563.59&181.67&431&18.05 
\\2942.8595&2943.6359&0.78&5.91&0.30&-3.48E+04&1.78E+03&5.12E+07&2.61E+06&6.416&2.848&5.323&0.033&191.55&96.82&424&15.96 
\\2943.6373&2944.4137&0.78&-4.94&0.35&2.91E+04&2.05E+03&-4.28E+07&3.02E+06&10.966&2.929&5.239&0.021&439.01&60.73&309&14.57 
\\2944.4151&2945.1901&0.77&2.41&1.49&-1.42E+04&8.78E+03&2.09E+07&1.29E+07&10.673&2.979&5.297&0.021&270.53&61.78&431&16.04 
\\2945.1914&2945.9678&0.78&8.80&1.04&-5.19E+04&6.12E+03&7.64E+07&9.02E+06&7.049&3.309&5.454&0.034&-194.04&99.28&430&18.14 
\\2945.9692&2946.7456&0.78&-2.73&1.11&1.61E+04&6.53E+03&-2.37E+07&9.62E+06&5.965&2.963&5.565&0.036&-521.14&106.68&430&16.44 
\\2946.7470&2947.5220&0.77&8.56&2.02&-5.04E+04&1.19E+04&7.43E+07&1.75E+07&7.569&3.089&5.394&0.029&-17.17&86.84&431&17.45 
\\2947.5233&2948.2997&0.78&-9.72&0.96&5.73E+04&5.66E+03&-8.44E+07&8.35E+06&11.689&2.446&5.280&0.016&317.16&46.79&430&15.93 
\\2948.3011&2949.0775&0.78&2.53&1.44&-1.49E+04&8.50E+03&2.20E+07&1.25E+07&9.862&3.094&5.333&0.023&162.11&68.40&430&17.09 
\\2949.0789&2949.7469&0.67&9.56&1.07&-5.64E+04&6.33E+03&8.32E+07&9.33E+06&7.684&3.774&5.395&0.035&-19.06&103.75&385&17.64 
\\2955.0440&2956.0731&1.03&-2.62&1.00&1.55E+04&5.93E+03&-2.29E+07&8.77E+06&12.264&2.403&5.368&0.014&61.09&40.79&581&24.63 
\\2956.0745&2956.8495&0.77&-15.37&0.77&9.09E+04&4.54E+03&-1.34E+08&6.72E+06&16.820&2.900&5.331&0.013&170.16&37.87&307&17.42 
\\2956.8509&2957.6273&0.78&0.67&1.41&-3.99E+03&8.33E+03&5.89E+06&1.23E+07&15.218&3.003&5.325&0.015&186.27&43.22&430&18.42 
\\2957.6286&2958.4050&0.78&7.40&2.37&-4.38E+04&1.40E+04&6.47E+07&2.07E+07&15.357&3.801&5.311&0.018&227.97&53.93&430&22.37 
\\2958.4064&2959.1814&0.77&22.57&1.21&-1.34E+05&7.15E+03&1.98E+08&1.06E+07&9.905&3.263&5.500&0.024&-329.40&70.43&431&20.97 
\\2959.1828&2959.9591&0.78&237.19&1.43&-1.40E+06&8.47E+03&2.08E+09&1.25E+07&-2.830&4.089&5.279&0.108&324.63&319.74&430&28.05 
\\2959.9605&2960.7369&0.78&-1.85&1.80&1.09E+04&1.06E+04&-1.61E+07&1.57E+07&5.804&2.860&5.643&0.055&2786.40&162.34&430&44.58 
\\2960.7383&2961.5133&0.77&-187.89&0.49&1.11E+06&2.90E+03&-1.65E+09&4.29E+06&12.141&9.469&5.763&0.055&-1110.48&162.45&431&162.90 
\\2961.5147&2962.5730&1.06&0.14&0.69&-8.38E+02&4.08E+03&1.26E+06&6.04E+06&-13.302&4.825&5.235&0.039&-0.44&0.04&602&175.33 
\\
	\bottomrule
\end{longtable}
\footnotesize{ $^a$ BJD - 2457000;
	$^b$ Degree of freedom of the fit;
	$^c$ Fitted reduced chi-square.}	
\renewcommand\thetable{A2}
\setlength{\tabcolsep}{0.1pt}
\begin{longtable}{ccccccccccccccccc}

	\caption{Eclipse parameters and O-C analysis of the quiescence of HS 2325+8205.}\label{tab:A2}\\

	
	\toprule
	Minima-BJD&err&Minima-flux&err&Eclipse depth&E&O-C&Minima-BJD&err&Minima-flux&err&Eclipse depth&E&O-C\\

[BJD - 2457000]&[d]&[e/s]&[e/s]&[e/s]& &[d]&[BJD - 2457000]&[d]&[e/s]&[e/s]&[e/s]& &[d]\\
	\midrule
	\endfirsthead
	\multicolumn{14}{c}
	{{\bfseries{Table \thetable\ }continued from previous page}} \\
	\toprule
	Minima-BJD&err&Minima-flux&err&Eclipse depth&E&O-C&Minima-BJD&err&Minima-flux&err&Eclipse depth&E&O-C\\
	
	[BJD - 2457000]&[d]&[e/s]&[e/s]&[e/s]& &[d]&[BJD - 2457000]&[d]&[e/s]&[e/s]&[e/s]& &[d]\\
	\midrule
	\endhead
	\midrule
	\multicolumn{14}{c}
	{{ }} \\
	\endfoot
	\endlastfoot
2748.04965&0.00047&-21.96&3.43&18.69&4911&-0.00451&2923.14741&0.00054&-13.06&1.23&12.48&5812&-0.00236 \\
2748.24578&0.00097&-16.84&3.81&15.32&4912&-0.00271&2923.72959&0.00068&-11.23&1.74&11.37&5815&-0.00318 \\
2748.43980&0.00077&-18.10&3.87&17.89&4913&-0.00303&2923.92478&0.00166&-9.97&3.11&8.87&5816&-0.00233 \\
2748.63278&0.00067&-20.12&3.45&19.41&4914&-0.00438&2924.11897&0.00128&-11.00&2.49&9.63&5817&-0.00247 \\
2748.83026&0.00157&-17.29&3.47&17.20&4915&-0.00124&2924.31316&0.00085&-11.86&3.24&10.88&5818&-0.00262 \\
2749.02389&0.00101&-17.22&3.58&16.16&4916&-0.00194&2924.50516&0.00088&-11.86&2.29&10.56&5819&-0.00495 \\
2749.21852&0.00080&-18.97&2.77&18.66&4917&-0.00165&2924.70201&0.00073&-13.30&1.68&12.32&5820&-0.00243 \\
2749.41425&0.00107&-25.02&3.86&23.07&4918&-0.00025&2924.89448&0.00073&-12.94&1.65&12.51&5821&-0.00430 \\
2749.60627&0.00057&-25.82&2.99&22.76&4919&-0.00257&2925.09088&0.00156&-10.41&1.61&9.29&5822&-0.00223 \\
2749.80051&0.00082&-24.11&3.67&23.67&4920&-0.00266&2925.28401&0.00094&-9.64&1.71&8.70&5823&-0.00344 \\
2749.99575&0.00103&-21.31&3.80&21.68&4921&-0.00175&2925.47658&0.00106&-7.19&1.63&7.19&5824&-0.00520 \\
2750.18884&0.00065&-19.97&3.31&19.95&4922&-0.00300&2925.67411&0.00195&-4.45&1.48&3.15&5825&-0.00201 \\
2750.38168&0.00047&-22.28&3.03&22.40&4923&-0.00449&2925.86824&0.00114&-5.33&1.38&4.13&5826&-0.00221 \\
2750.57743&0.00099&-17.82&3.61&14.40&4924&-0.00308&2926.06046&0.00081&-6.66&1.76&6.25&5827&-0.00433 \\
2750.77369&0.00081&-15.51&2.63&13.96&4925&-0.00115&2926.25977&0.00228&-7.24&1.70&6.08&5828&0.00065 \\
2750.96527&0.00057&-20.75&2.83&19.89&4926&-0.00391&2926.45165&0.00119&-8.91&1.47&8.69&5829&-0.00181 \\
2751.16133&0.00101&-20.17&3.96&19.96&4927&-0.00218&2926.64694&0.00061&-10.85&1.33&9.42&5830&-0.00085 \\
2751.35439&0.00118&-17.21&3.20&17.54&4928&-0.00346&2926.84007&0.00059&-11.38&1.41&11.01&5831&-0.00206 \\
2751.54840&0.00215&-17.26&3.46&14.26&4929&-0.00378&2927.03478&0.00060&-12.06&1.55&11.33&5832&-0.00168 \\
2751.74390&0.00082&-16.32&2.54&14.61&4930&-0.00262&2927.22958&0.00060&-13.02&1.62&12.08&5833&-0.00122 \\
2751.93576&0.00085&-15.55&3.26&15.04&4931&-0.00509&2927.42332&0.00055&-15.66&1.44&14.63&5834&-0.00181 \\
2752.13182&0.00067&-14.87&2.83&14.10&4932&-0.00337&2927.61774&0.00054&-15.24&1.57&14.18&5835&-0.00173 \\
2752.32675&0.00085&-17.41&3.38&15.11&4933&-0.00277&2927.81199&0.00048&-15.35&1.54&15.37&5836&-0.00181 \\
2752.52385&0.00084&-17.37&2.82&16.58&4934&-0.00001&2928.00528&0.00043&-14.52&1.26&13.93&5837&-0.00286 \\
2752.71458&0.00066&-19.48&3.55&19.24&4935&-0.00361&2928.20064&0.00047&-11.25&1.28&10.45&5838&-0.00183 \\
2752.91075&0.00116&-16.10&3.73&15.84&4936&-0.00178&2928.39362&0.00046&-9.96&1.44&9.50&5839&-0.00319 \\
2753.10515&0.00080&-17.75&3.29&16.00&4937&-0.00171&2928.58864&0.00051&-11.34&1.59&10.24&5840&-0.00250 \\
2753.29912&0.00101&-21.72&3.38&18.44&4938&-0.00208&2928.78288&0.00050&-10.53&1.47&9.29&5841&-0.00259 \\
2753.49289&0.00056&-24.35&2.88&22.97&4939&-0.00264&2928.97731&0.00048&-12.66&1.62&11.83&5842&-0.00250 \\
2753.68608&0.00080&-23.68&3.48&22.98&4940&-0.00379&2929.17042&0.00048&-14.32&1.47&13.96&5843&-0.00372 \\
2753.88198&0.00067&-21.90&2.85&21.66&4941&-0.00222&2929.36408&0.00054&-14.62&1.56&14.56&5844&-0.00440 \\
2754.07636&0.00054&-22.76&2.99&22.43&4942&-0.00217&2929.55701&0.00064&-13.58&1.38&12.31&5845&-0.00580 \\
2754.26920&0.00059&-21.67&3.48&20.81&4943&-0.00367&2929.75128&0.00071&-12.99&1.21&11.31&5846&-0.00587 \\
2754.46355&0.00041&-21.24&2.59&18.02&4944&-0.00365&2929.94547&0.00163&-7.52&1.51&6.70&5847&-0.00601 \\
2754.65800&0.00054&-18.90&2.97&17.74&4945&-0.00354&2930.14159&0.00042&-8.36&1.01&7.05&5848&-0.00423 \\
2754.85240&0.00042&-21.68&1.94&19.93&4946&-0.00347&2930.53097&0.00065&-6.34&1.08&5.36&5850&-0.00352 \\
2755.04731&0.00064&-23.29&3.09&22.76&4947&-0.00290&2930.72632&0.00087&-6.66&1.45&7.00&5851&-0.00250 \\
2755.24130&0.00097&-22.44&3.10&22.08&4948&-0.00324&2930.92258&0.00093&-6.67&1.38&6.03&5852&-0.00058 \\
2755.43578&0.00094&-23.75&3.16&21.46&4949&-0.00310&2931.11665&0.00084&-8.07&1.17&7.28&5853&-0.00084 \\
2755.62961&0.00129&-20.05&2.41&16.91&4950&-0.00360&2931.31084&0.00099&-10.18&1.64&8.84&5854&-0.00099 \\
2755.82614&0.00156&-12.79&2.69&12.37&4951&-0.00141&2931.50517&0.00059&-11.27&1.29&10.09&5855&-0.00099 \\
2756.98983&0.00066&-20.71&2.75&18.27&4957&-0.00373&2931.69948&0.00101&-10.79&1.95&9.80&5856&-0.00102 \\
2757.18498&0.00070&-19.68&2.92&17.68&4958&-0.00291&2931.89287&0.00040&-12.44&1.12&11.47&5857&-0.00196 \\
2757.38224&0.00102&-16.33&2.92&15.26&4959&0.00001&2932.08672&0.00050&-15.55&1.43&14.32&5858&-0.00245 \\
2757.57369&0.00073&-20.34&3.10&19.23&4960&-0.00287&2932.28175&0.00055&-15.10&1.50&14.77&5859&-0.00175 \\
2757.76961&0.00069&-20.14&2.76&21.02&4961&-0.00128&2932.47518&0.00047&-14.46&1.38&14.34&5860&-0.00266 \\
2757.96156&0.00053&-24.15&2.96&25.25&4962&-0.00367&2932.66928&0.00050&-12.23&1.20&12.26&5861&-0.00289 \\
2758.15652&0.00039&-29.12&2.83&29.82&4963&-0.00304&2932.86454&0.00058&-9.70&1.53&9.14&5862&-0.00196 \\
2758.35119&0.00083&-19.23&3.27&16.97&4964&-0.00271&2933.05832&0.00048&-11.03&1.63&9.78&5863&-0.00252 \\
2758.54544&0.00088&-18.90&3.85&16.44&4965&-0.00279&2933.25213&0.00040&-11.90&1.52&10.88&5864&-0.00304 \\
2758.73828&0.00046&-22.37&3.51&19.61&4966&-0.00429&2933.44533&0.00035&-10.43&1.06&9.12&5865&-0.00418 \\
2758.93307&0.00054&-23.38&3.82&22.26&4967&-0.00383&2933.64018&0.00060&-10.45&1.64&9.87&5866&-0.00366 \\
2759.12866&0.00088&-21.11&4.06&20.26&4968&-0.00258&2933.83523&0.00061&-12.68&1.56&12.29&5867&-0.00295 \\
2759.32138&0.00055&-25.28&3.13&24.27&4969&-0.00419&2934.03044&0.00056&-13.21&1.18&12.55&5868&-0.00207 \\
2759.51678&0.00104&-26.15&3.24&24.26&4970&-0.00313&2934.22300&0.00053&-15.49&1.23&14.48&5869&-0.00385 \\
2759.70951&0.00084&-22.59&3.17&21.82&4971&-0.00473&2934.41680&0.00066&-14.71&1.54&14.12&5870&-0.00438 \\
2759.90342&0.00036&-22.54&2.18&22.37&4972&-0.00516&2942.38812&0.00241&-10.71&3.65&9.83&5911&-0.00079 \\
2760.10069&0.00184&-10.95&2.59&9.01&4973&-0.00222&2942.58147&0.00276&-11.88&3.57&11.36&5912&-0.00177 \\
2760.29357&0.00072&-17.07&3.03&14.09&4974&-0.00368&2942.77686&0.00151&-16.43&3.07&13.81&5913&-0.00072 \\
2760.48695&0.00045&-15.61&2.53&14.02&4975&-0.00463&2942.97188&0.00089&-20.82&2.59&17.89&5914&-0.00003 \\
2760.68196&0.00124&-10.74&3.62&9.95&4976&-0.00396&2943.16433&0.00118&-18.89&3.60&17.46&5915&-0.00192 \\
2760.87991&0.00141&-13.05&3.63&12.35&4977&-0.00034&2943.35972&0.00143&-18.34&3.42&17.72&5916&-0.00086 \\
2761.07346&0.00096&-17.00&3.30&16.75&4978&-0.00113&2943.55289&0.00111&-18.06&3.99&17.76&5917&-0.00203 \\
2761.26718&0.00087&-15.95&3.00&16.05&4979&-0.00174&2943.93918&0.00216&-21.25&3.52&19.23&5919&-0.00441 \\
2761.46157&0.00084&-17.85&3.23&15.61&4980&-0.00169&2944.13654&0.00075&-24.03&3.81&23.69&5920&-0.00138 \\
2761.65772&0.00113&-19.39&3.83&17.70&4981&0.00013&2944.32897&0.00079&-23.51&3.67&21.95&5921&-0.00328 \\
2761.85170&0.00091&-26.31&4.09&22.02&4982&-0.00022&2944.52452&0.00064&-17.69&2.28&16.98&5922&-0.00207 \\
2762.04507&0.00103&-21.80&3.23&18.44&4983&-0.00119&2944.71907&0.00122&-16.05&3.36&14.98&5923&-0.00185 \\
2762.23769&0.00049&-22.56&2.30&22.74&4984&-0.00290&2944.91234&0.00109&-16.93&2.98&14.79&5924&-0.00292 \\
2762.43284&0.00052&-22.68&2.25&24.07&4985&-0.00209&2945.10793&0.00094&-18.84&2.83&17.62&5925&-0.00166 \\
2762.62501&0.00058&-24.96&3.96&26.38&4986&-0.00425&2945.29957&0.00077&-20.25&2.89&18.59&5926&-0.00436 \\
2762.82078&0.00040&-25.70&2.91&25.07&4987&-0.00282&2945.49446&0.00120&-17.25&3.52&16.46&5927&-0.00380 \\
2763.01470&0.00055&-19.72&3.16&18.80&4988&-0.00323&2945.68867&0.00089&-17.48&3.33&17.29&5928&-0.00393 \\
2763.20932&0.00045&-19.95&3.15&16.88&4989&-0.00295&2945.88274&0.00060&-16.75&2.79&15.84&5929&-0.00419 \\
2763.40369&0.00044&-17.67&2.60&15.87&4990&-0.00291&2946.07712&0.00121&-9.94&2.73&8.98&5930&-0.00415 \\
2763.59706&0.00055&-25.52&3.96&21.93&4991&-0.00388&2946.27196&0.00098&-11.59&2.52&11.77&5931&-0.00364 \\
2763.79231&0.00064&-20.11&2.89&20.57&4992&-0.00296&2946.47007&0.00159&-12.73&3.24&10.85&5932&0.00013 \\
2763.98505&0.00053&-25.06&2.52&24.17&4993&-0.00456&2946.66503&0.00149&-17.82&4.35&16.01&5933&0.00076 \\
2764.18017&0.00077&-23.64&2.66&22.31&4994&-0.00377&2946.85531&0.00082&-17.34&3.18&15.64&5934&-0.00330 \\
2916.15166&0.00067&-12.98&1.63&11.73&5776&-0.00206&2947.05224&0.00090&-17.34&2.77&15.16&5935&-0.00070 \\
2916.34539&0.00073&-12.65&1.60&10.85&5777&-0.00266&2947.24549&0.00120&-17.23&4.31&16.34&5936&-0.00179 \\
2916.54014&0.00062&-12.77&1.35&12.26&5778&-0.00225&2947.44141&0.00088&-19.69&3.23&18.84&5937&-0.00020 \\
2916.73133&0.00045&-13.17&1.31&13.81&5779&-0.00539&2947.63420&0.00086&-25.62&4.15&23.54&5938&-0.00175 \\
2917.12045&0.00062&-11.17&1.41&10.40&5781&-0.00494&2947.82812&0.00092&-22.07&3.78&21.37&5939&-0.00216 \\
2917.31567&0.00061&-9.41&1.28&8.51&5782&-0.00405&2948.02115&0.00078&-21.93&3.52&20.57&5940&-0.00347 \\
2917.51042&0.00033&-9.74&1.06&8.94&5783&-0.00364&2948.21625&0.00067&-21.82&3.41&19.19&5941&-0.00270 \\
2917.70551&0.00075&-9.05&1.59&7.94&5784&-0.00288&2948.41086&0.00083&-19.16&3.79&18.09&5942&-0.00242 \\
2917.90111&0.00092&-9.46&1.47&7.99&5785&-0.00162&2948.60521&0.00114&-16.55&4.07&14.84&5943&-0.00241 \\
2918.09541&0.00074&-10.47&1.39&9.41&5786&-0.00165&2948.79783&0.00061&-21.49&3.96&19.04&5944&-0.00412 \\
2918.28972&0.00069&-9.84&1.17&8.93&5787&-0.00168&2948.99211&0.00075&-18.17&2.78&16.30&5945&-0.00418 \\
2918.48471&0.00105&-9.99&1.73&9.35&5788&-0.00102&2949.18803&0.00127&-17.02&3.47&16.20&5946&-0.00259 \\
2918.67809&0.00090&-11.73&1.44&11.34&5789&-0.00198&2949.38115&0.00132&-17.67&3.79&17.28&5947&-0.00381 \\
2918.87212&0.00050&-14.61&1.26&13.51&5790&-0.00228&2949.57569&0.00157&-15.43&3.40&15.45&5948&-0.00360 \\
2919.06629&0.00059&-13.86&1.34&12.58&5791&-0.00245&2955.21221&0.00077&-21.81&3.93&21.17&5977&-0.00279 \\
2919.26150&0.00055&-13.15&1.36&13.16&5792&-0.00157&2955.40790&0.00118&-20.50&3.40&18.05&5978&-0.00144 \\
2919.45384&0.00054&-12.64&1.53&12.14&5793&-0.00357&2955.60228&0.00088&-20.76&3.79&20.28&5979&-0.00139 \\
2919.64851&0.00049&-12.41&1.55&12.02&5794&-0.00323&2955.79743&0.00120&-21.04&3.36&21.11&5980&-0.00058 \\
2919.84409&0.00050&-11.23&1.23&10.42&5795&-0.00199&2955.98975&0.00065&-24.62&3.28&24.70&5981&-0.00259 \\
2920.03730&0.00066&-11.08&1.61&9.78&5796&-0.00311&2956.18381&0.00099&-25.88&4.01&24.57&5982&-0.00286 \\
2920.23160&0.00051&-10.69&1.16&9.82&5797&-0.00315&2956.57246&0.00047&-28.39&3.46&27.33&5984&-0.00288 \\
2920.42589&0.00065&-11.04&1.66&10.29&5798&-0.00319&2956.76733&0.00047&-22.82&2.86&21.81&5985&-0.00235 \\
2920.61895&0.00052&-11.48&1.48&11.42&5799&-0.00447&2956.96045&0.00044&-20.98&3.17&19.62&5986&-0.00356 \\
2920.81362&0.00059&-11.33&1.44&11.29&5800&-0.00413&2957.15473&0.00050&-18.70&3.40&17.09&5987&-0.00362 \\
2921.00711&0.00045&-11.31&0.98&10.21&5801&-0.00497&2957.35082&0.00105&-16.06&4.34&15.10&5988&-0.00186 \\
2921.20221&0.00066&-9.90&1.27&9.18&5802&-0.00421&2957.54381&0.00122&-20.99&5.40&19.16&5989&-0.00321 \\
2921.39641&0.00195&-5.88&0.98&4.11&5803&-0.00434&2957.73702&0.00057&-24.35&3.20&23.83&5990&-0.00433 \\
2921.59025&0.00180&-5.96&0.99&4.69&5804&-0.00484&2957.93126&0.00069&-27.98&3.53&27.49&5991&-0.00443 \\
2921.78913&0.00332&-6.56&1.45&5.44&5805&-0.00029&2958.12642&0.00067&-26.39&2.63&25.40&5992&-0.00360 \\
2921.98264&0.00093&-8.64&1.38&8.44&5806&-0.00112&2958.31989&0.00060&-25.92&2.44&23.03&5993&-0.00447 \\
2922.17629&0.00068&-10.44&1.46&9.37&5807&-0.00180&2958.51565&0.00099&-20.34&2.79&17.75&5994&-0.00304 \\
2922.37077&0.00068&-10.57&1.50&9.43&5808&-0.00166&2958.71039&0.00117&-14.95&3.01&12.61&5995&-0.00264 \\
2922.56452&0.00090&-9.95&1.70&9.43&5809&-0.00224&2958.90284&0.00085&-13.15&2.89&11.13&5996&-0.00452 \\
2922.75983&0.00065&-12.88&1.46&11.00&5810&-0.00127&2959.09804&0.00084&-15.99&2.97&15.44&5997&-0.00366 \\
2922.95440&0.00070&-11.33&1.25&10.63&5811&-0.00103&-&-&-&-&-&-&- \\

	\bottomrule
\end{longtable}

\end{document}